\documentclass[aps,showpacs,prb,twocolumn,superscriptaddress,floatfix,longbibliography]{revtex4-2}

\usepackage{graphicx,bm,amssymb,amsmath,dcolumn,hyperref}
\usepackage{multirow}
\usepackage[shortlabels]{enumitem}
\usepackage{color}
\usepackage{subfigure}  
\usepackage{epstopdf}
\usepackage{bbm}
\usepackage{graphicx}
\usepackage{hyperref}
\usepackage{threeparttable}
\usepackage{amsthm}
\usepackage{mathtools}
\usepackage{color}
\usepackage{tikz}
\usepackage{float}
\usepackage{empheq}
\usepackage{algorithm}
\usepackage{algpseudocode}
\usepackage{array}
\usepackage{tabularx}

\begin{document}
\title{The Nonclassical Regime of the Two-dimensional Long-range XY Model:\\ a Comprehensive Monte Carlo Study}
	
\author{Dingyun Yao}
\thanks{These two authors contributed equally to this work}
\affiliation{Hefei National Research Center for Physical Sciences at the Microscale and School of Physical Sciences, University of Science and Technology of China, Hefei 230026, China}

\author{Tianning Xiao}
\thanks{These two authors contributed equally to this work}
\affiliation{Hefei National Research Center for Physical Sciences at the Microscale and School of Physical Sciences, University of Science and Technology of China, Hefei 230026, China}

\author{Chao Zhang}
\affiliation{Hefei National Research Center for Physical Sciences at the Microscale and School of Physical Sciences, University of Science and Technology of China, Hefei 230026, China}
\affiliation{Department of Physics, Anhui Normal University, Wuhu, Anhui 241000, China}

\author{Youjin Deng}
\email{yjdeng@ustc.edu.cn}
\affiliation{Hefei National Research Center for Physical Sciences at the Microscale and School of Physical Sciences, University of Science and Technology of China, Hefei 230026, China}
\affiliation{Hefei National Laboratory, University of Science and Technology of China, Hefei 230088, China}
\affiliation{Shanghai Research Center for Quantum Science and CAS Center for Excellence in Quantum Information and Quantum Physics, University of Science and Technology of China, Shanghai 201315, China}

\author{Zhijie Fan}
\email{zfanac@ustc.edu.cn}
\affiliation{Hefei National Research Center for Physical Sciences at the Microscale and School of Physical Sciences, University of Science and Technology of China, Hefei 230026, China}
\affiliation{Hefei National Laboratory, University of Science and Technology of China, Hefei 230088, China}
\affiliation{Shanghai Research Center for Quantum Science and CAS Center for Excellence in Quantum Information and Quantum Physics, University of Science and Technology of China, Shanghai 201315, China}

\begin{abstract}
{The two-dimensional (2D) XY model plays a crucial role in statistical and condensed matter physics. With the introduction of long-range interactions, the system exhibits a richer set of physical phenomena and a crossover between non-classical and short-range universality classes.}
In this work, we investigate the 2D XY model with algebraically decaying interactions $\sim 1/r^{2+\sigma}$, and provide a comprehensive numerical analysis of its thermodynamic properties. We demonstrate that for $\sigma \leq 2$, the system undergoes a second-order phase transition into a ferromagnetic phase characterized by the emergence of long-range order. In the low-temperature phase, due to the presence of the Goldstone mode, the correlation function saturates to a non-zero constant in the form of a power law for $\sigma < 2$, with decaying exponent $2-\sigma$, and in the form of the inverse logarithm of distance for $\sigma=2$.  {Moreover, the critical points and exponents are also determined for various $\sigma$. We provide compelling evidence that the crossover between non-classical and short-range regimes occurs at $\sigma=2$. This work presents a detailed account of the simulation methodology, extensive numerical data, and new insights into the physics of long-range interacting systems.}
\end{abstract}

\maketitle

\section{Introduction}

Long-range (LR) interactions are prevalent in various natural phenomena, such as gravitational interactions between celestial bodies and electromagnetic interactions between particles. A key characteristic of these interactions is that their strength decays with distance according to a power law. Introducing LR interactions into complex many-body systems can significantly influence their phase transition properties, leading to a richer array of physical phenomena. The first study of LR interactions dates back to 1969, when Dyson investigated their effects in a one-dimensional (1D) Ising chain by introducing the interaction term $J_{ij} = |i-j|^{-(1+\sigma)}$ \cite{Dyson1969}. Dyson's work revealed that, unlike the original Ising model, which exhibited no phase transitions, the LR interactions induce a second-order phase transition at finite temperatures for $0 < \sigma < 1$. This finding highlights the profound impact of LR interactions on phase transitions, demonstrating that they can give rise to novel and intriguing physical behaviors in many-body systems.

Building upon this foundational work, the Ising model can be extended to the general $O(n)$ spin model with LR interactions, where the interaction decays as $r_{ij}^{-(d+\sigma)}$. In this context, $r_{ij}$ represents the distance between site $i$ and $j$, and $d$ is the spatial dimension. This extension provides a broader framework for understanding the effects of LR interactions in various many-body systems.

The LR $O(n)$ spin model was first thoroughly studied by Fisher using a renormalization group (RG) approach~\cite{Fisher1972}. He categorized the system's behavior into three regimes based on the parameter $\sigma$: (1) $0 < \sigma < \frac{d}{2}$ (classical regime): within this regime, the critical behavior of the system is governed by the Gaussian fixed point; (2) $\frac{d}{2} < \sigma < 2$ (nonclassical regime): Gaussian fixed points cease to be stable, and the critical exponents vary with $\sigma$, with the critical exponent $\eta$ posited to precisely retain its value within the mean-field regime, namely $\eta = 2 - \sigma$; (3) $\sigma > 2$ (short-range (SR) regime): in this interval, LR terms in the Hamiltonian become irrelevant, leading to the recovery of SR behavior. Regimes (1) and (3) have gained widespread acceptance, whereas regime (2) presents some issues. According to the regime (2), the exponent $\eta$ might exhibit a sudden jump from 0 to $\eta_{\text{SR}}$ at $\sigma = 2$, where $\eta_{\text{SR}}$ denotes the $\eta$ value under short-range interactions. Later, Sak proposed an alternative framework, known as Sak's criterion, where the boundary between the nonclassical and short-range regimes shifts from 2 to $\sigma_* = 2 - \eta_{\text{SR}}$. In this revised framework, $\eta = 2 - \sigma$ is retained in the interval $1 < \sigma < \sigma_*$, and $\eta = \eta_{\text{SR}}$ for $\sigma > \sigma_*$, thus simplifying $\eta$ to $\eta = \max(2 - \sigma, \eta_{\text{SR}})$, eliminating the abrupt change in the exponent $\eta$. Sak's criterion has been supported by a series of theoretical and numerical works~\cite{Honkonen_1989, Honkonen_1990, defenu2015, Luijten2002, angelini2014, horita2017}. 

However, later Monte Carlo (MC) results with higher precision and subsequent theoretical work suggest a different picture, where $\eta$ transitions smoothly from $2 - \sigma$ near $\sigma = 1$ to $\eta_{\text{SR}}$ at $\sigma = 2$, and the threshold $\sigma_*$ returns to $2$~\cite{picco2012, blanchard2013}. Additionally, another MC study of a percolation model with LR probabilities appears to support this picture~\cite{Grassberger2013}. Later, another numerical study indicated the possible existence of a double power-law in the correlation function near $\sigma = 2 - \eta_{\text{SR}}$, attributing this deviation to an underestimation of finite-size corrections~\cite{angelini2014}.

Overall, Sak's criterion has achieved considerable recognition, but controversies persist. Most numerical validations have been focused on the two-dimensional (2D) LR Ising model, with limited exploration of LR $O(n)$ spin models for $n \geqslant 2$. For instance, Ref.~\cite{berganza2013} studied the XY model on a 2D complex topology, which is expected to belong to the same universality class as the 2D LR XY model. Similarly, Ref.~\cite{Zhao2023} investigated the 2D LR quantum Heisenberg model at finite temperatures. However, these studies did not focus on determining the threshold point $\sigma_*$, and their simulations were limited to relatively small system sizes, with a maximum length of only 256. Therefore, there is a pressing need for a more detailed numerical study of these models.

Apart from its implications for Sak’s criterion, the phase diagram of the 2D LR XY model exhibits novel characteristics. Although the Mermin–Wagner theorem forbids true long-range order (LRO) in a 2D XY model with finite-range coupling~\cite{mermin1966, mermin_absence_1967}, the system undergoes a BKT transition with an exponent $\eta_{\mathrm {SR}} = 1/4$, entering a low-temperature (low-T) quasi-long-range ordered (QLRO) phase~\cite{kosterlitz2017}. In contrast, in the mean-field regime $\sigma<\frac{d}{2}$, the XY model exhibits an ordinary second-order transition to a LRO phase. Thus, one would expect that in the nonclassical regime, there exists a point where the transition type changes drastically from BKT to second-order. The nature of this crossover regime remains a key open question. The Mermin–Wagner theorem applies under the condition that the second moment of the interaction $\sum_{\vec{r}} r^2 |J(\vec{r})|$ is finite~\cite{mermin1966, mermin_absence_1967}. For algebraically decaying interactions $J(r) \sim 1/r^{2+\sigma}$, this condition is violated when $\sigma \leq 2$ due to the divergence of the sum. This suggests that the theorem does not exclude
the possibility of LRO for $\sigma \le 2$. In 2001, Bruno provided a more stringent generalization of the Mermin-Wagner theorem for one- and two-dimensional LR XY and Heisenberg systems, stating that an LRO phase is allowed at sufficiently low temperatures for $\sigma<2$ but excluded for $\sigma\geq2$~\cite{PhysRevLett.87.137203}. Moreover, a numerical study has shown that the 2D LR diluted XY model displays spontaneous magnetization at low temperatures for $\sigma < 2$~\cite{Cescatti2019}. This suggests that if Sak's criterion holds for the 2D LR XY model, then within the interval $1.75 < \sigma \le 2$, the system should exhibit both a BKT transition and a low-T ferromagnetic phase. This scenario is particularly intriguing -- recent works~\cite{Giachetti2021, Giachetti2022} have proposed a phase diagram where the LR XY system exhibits two phase transitions for $1.75 < \sigma < 2$: first a BKT transition into a QLRO phase, followed by another transition into a ferromagnetic phase.

\begin{figure}[t]
    \centering
    \includegraphics[width=0.9\linewidth]{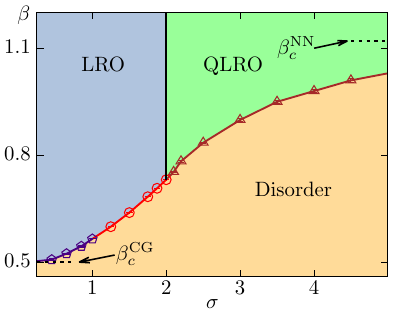}
    \caption{The phase diagram of the two-dimensional long-range XY model reveals distinct behaviors depending on the interaction range $\sigma$ and the inverse temperature $\beta=1/T$. In the short-range regime ($\sigma > 2$), the system exhibits BKT transitions (brown line) into a QLRO phase. In the nonclassical regime ($1 < \sigma \leq 2$), the system undergoes a continuous phase transition (red line) into a long-range ordered phase. Lastly, in the classical regime ($\sigma \leq 1$), the transition (purple lines) is governed by the Gaussian fix point. The symbols $\beta_c^{\text{CG}}$ and $\beta_c^{\text{NN}}$ denote the critical inverse temperatures for the complete-graph and NN interaction limits, respectively.}
    \label{PD}
\end{figure}

This paper provides a comprehensive study of the 2D LR XY model, offering new insights into its critical behavior and phase diagram. We perform large-scale Monte Carlo (MC) simulations with the largest linear system size up to $L = 8192$. Instead of further refining the estimate of critical exponents in the nonclassical regime, we focus on both the low- and high-temperature (high-T) properties of the model, and a brief summary of the main results is given in Ref.~\cite{xiao2024}. The phase diagram for the 2D LR XY model is shown in Figure~\ref{PD}. For $\sigma > 2$, the system displays SR behavior: a BKT transition to QLRO for the XY model. For $\sigma \leq 2$, the system exhibits a second-order phase transition, with critical exponents varying according to $\sigma$. In this work, we present a series of evidence that strongly indicates that the threshold between LR and SR universality is at $\sigma_* = 2$ for the 2D LR XY model, which is inconsistent with Sak's scenario and the proposed phase diagram in Ref.~\cite{Giachetti2021, Giachetti2022}.

1. The second-moment correlation length ratio $\xi/L$ (will be defined in the next section) is a powerful tool for identifying the transition type and the critical point. Specifically, for a second-order transition, $\xi/L$ curves for different $L$s intersect at the critical point and diverge in the low-T phases as $L$ increases. In contrast, for a BKT transition, $\xi/L$ curves for different $L$s converge to a universal function in the QLRO phase. Our results show that for $\sigma \le 2$, the scaling behavior of $\xi/L$ is consistent with the second-order transition, while for $\sigma > 2$, typical BKT physics is observed.

2. In the whole low-T phases, we demonstrate that, for $\sigma\le2$, the system exhibits long-range ferromagnetic order in the thermodynamic limit, while for $\sigma>2$, the system enters a QLRO phase. In addition, the spontaneous symmetry breaking for $\sigma < 2$ leads to Goldstone mode excitations, which give rise to a correlation function that decays algebraically to a constant, as $g(r)\sim r^{-\eta_\ell} + g_0$. Here, the anomalous dimension $\eta_\ell = 2-\sigma$ is theoretically derived and confirmed numerically. Moreover, for the marginal case of $\sigma = 2$, where theoretical derivation is absent, logarithmic behavior is observed in the low-T phase, i.e. $g(r)\sim 1/\ln r +g_0$.

3. In the high-T phases, as $T$ decreases to the critical point $T_c$, the growth behavior of $\xi$ for $\sigma \le 2$ and $\sigma > 2$ are drastically different. For $\sigma > 2$, typical BKT behavior is observed, i.e., $\xi \sim \exp(b/\sqrt{t})$, where $t=(T-T_c)/T_c$ is the reduced temperature and $b$ is some non-universal constant. However, for $\sigma \le 2 $, the growth of $\xi$ increasingly deviates from the BKT behavior as $t\rightarrow0$. It asymptotically follows the power-law divergence characteristic of a second-order transition as $\xi \sim t^{-\nu}$, where $\nu$ is the critical exponent of the correlation length.

4. We carefully determine the critical temperature and critical exponents of the transitions for $\sigma \le 2$. The anomalous magnetic dimensions $\eta$ and the correlation-length exponent $\nu$ are continuous functions of $\sigma$ as long as $\sigma \leq 2$.

The rest of this paper is organized as follows. We first describe the model and present the enhanced MC algorithm in Section~\ref{sec::algorithm_and_observable}. The observables measured and their finite-size scaling analysis are also included. Sections~\ref{sec:results_phase_diagram}-\ref{sec:results_III} then present detailed results and analysis, including an overall explanation of the system at various $\sigma$, the low-T and high-T behaviors, and the critical exponents along the line of phase transitions. Finally, we summarize our findings in Section~\ref{sec:conclusion}.

\section{Model, Algorithm, and Observables}
\label{sec::algorithm_and_observable}
\subsection{Model}
We investigate the LR XY model on a two-dimensional square lattice with periodic boundary conditions (PBC), with the Hamiltonian as,
\begin{align}
    \mathcal{H} = - \sum_{i < j}^{N} J_{ij} \bm{S}_i \cdot \bm{S}_j,
    \label{eq_Hamiltonian}
\end{align}
where $\bm{S}_i$ is a two-component unit vector corresponding to the XY spin at the $i$th site, and $N=L\times L$ is the total number of spins. The summation runs over all spin pairs, and the total number of interaction terms is $N(N-1)/2$. The spins interact with each other via an algebraically decaying ferromagnetic coupling, 
\begin{align}
    J_{ij}=\frac{1}{\left|\vec{r}_i-\vec{r}_j\right|^{2+\sigma}},
    \label{couplingConstant}
\end{align}
where $\vec{r}_i$ is the coordinate of the $i$th site.
In the $\sigma \rightarrow \infty$ limit, the model reduces to the nearest-neighbor (NN) XY model, which enters a quasi-long-range order phase at low temperature through the celebrated BKT transition~\cite{kosterlitz2017}. In the $\sigma \rightarrow -2$ limit, this model becomes the XY model on a complete graph (CG), which corresponds to the infinite-dimensional lattice with PBC in the thermodynamic limit~\cite{aizenman1980,kirkpatrick2016,kirkpatrick2017}.
In this work, we focus on the $\sigma > 0$ regime, where the energy is strictly extensive, i.e., the energy density of the system remains finite in the thermodynamic limit.

To properly account for the long-range interactions across the periodic boundaries, the minimum-image convention is employed~\cite{frenkel2002,janke2019,agrawal2021}. The square lattice with PBC maps onto a torus, and spins interact via the shortest distance on the surface of the torus. Specifically, the distance between any two lattice sites in the $x$ or $y$ direction does not exceed $L/2$. 

In the simulation, we introduce a normalization constant $c(\sigma,N)$ which modifies Eq.~\eqref{couplingConstant} as 
\begin{equation}
    J_{ij}=\frac{c(\sigma,N)}{\left|\vec{r}_i-\vec{r}_j\right|^{2+\sigma}},
    \label{normalization}
\end{equation}
such that $\sum_j J_{ij} = 4$. With such normalization, when $\sigma\to \infty$, the LR model reduces to the standard XY model with nearest neighboring interactions; when $\sigma\to -2$, $c(\sigma, N) = 4/(N-1)$ also recovers the standard XY model on the finite complete graph.
Another convention to deal with the long-range interactions in finite systems is the Ewald summation technique, where the coupling constant $J_{ij}$ is modified to include the interactions of all periodic images of the system~\cite{Luijten2002, horita2017}. In the thermodynamic limit, this approach is equivalent to our normalization scheme, and the two only exhibit minor differences in finite systems, which quickly vanish as $L$ increases.

\subsection{Algorithm} \label{subsec:algorithm} 
Simulating long-range interactions using the Monte Carlo method presents two major challenges: critical slowing down near phase transitions and escalating computational costs with increasing system sizes. Critical slowing down occurs near the critical point, where successive samples become highly correlated, resulting in a substantial reduction in simulation efficiency as system size increases. Various update schemes have been developed to address this issue, such as cluster algorithms~\cite{swendsen1987,wolff1989}, direct-loop algorithms~\cite{syljuasen2002}, and worm algorithms~\cite{prokofev2001}. The second challenge stems from the nature of long-range interactions: each spin interacts with the other $N-1$ spins, so a proposed local MC update requires evaluating all $N-1$ interaction terms. This results in a \textit{computational complexity} $\mathcal{C}$ of $\mathcal{O}(N)$ for the LR system, in contrast to $\mathcal{O}(1)$ for the NN case. We define the computational complexity as the average number of operations required to update a single spin, serving as a measure of the time complexity of the MC algorithm. In practice, the CPU time per local update scales linearly with $\mathcal{C}$. Several methods have been proposed to mitigate this cost. For example, the worm algorithm combined with the diagrammatic Monte Carlo method expands the pairwise potential energy into diagrammatic contributions, making the computational complexity independent of the system size~\cite{boninsegni2006}. Efficient cluster algorithms tailored for the classical long-range Ising model have also been developed~\cite{Luijten1995, fukui2009}. More recently, the clock Monte Carlo method~\cite{michel2019, Fan2023}, based on the factorized Metropolis filter~\cite{michel2014}, reduces computational complexities of several classical LR models to $\mathcal{O}(1)$, thus offering a scalable and efficient solution to simulating large-scale LR systems.

We simulate the LR XY model using a variant of the Luijten-Bl\"ote (LB) algorithm~\cite{Luijten1995, Luijten1997}. We enhance the original LB algorithm with the clock Monte Carlo method and extend it to the $O(n)$ spin model. This approach employs the cluster updates scheme to address the issue of critical slowing down and adopts an efficient cluster construction strategy to reduce the computational complexity associated with long-range interactions. 

We begin with the standard Swendson-Wang algorithm for a long-range $O(n)$ spin model~\cite{swendsen1987}. A Monte Carlo sweep consists of two main steps: cluster construction and spin flipping. The cluster construction starts by randomly choosing a unit vector $\mathbf{S}_{\rm{ref}}$ in the $n$-dimensional space. Each spin $\mathbf{S}$ is projected onto this reference vector, yielding the parallel component $S^{\parallel} = \mathbf{S}\cdot\mathbf{S}_{\rm{ref}}$. Only this projection component can undergo spin flipping during the update process, while the remaining orthogonal components remain invariant. This is effectively an Ising-type update scheme with the coupling constant between $\boldsymbol{S}_i$,$\boldsymbol{S}_j$ as $J_{ij}S^{\parallel}_i S_j^{\parallel}$. Spin clusters are then formed by sequentially inspecting all pairs of spins (bonds). Each bond $(i,j)$ is independently activated with probability,
\begin{equation}
    \rho_{ij} = \left[1 - e^{-2 \beta J_{ij}S^{\parallel}_iS^{\parallel}_j}\right]^{+},
    \label{eq:bond_evaluation}
\end{equation}
with $[x]^+ = {\rm max}(0,x)$ and the inverse temperature $\beta=1/T$. 
After all the bonds have been examined, spins connected by active bonds form clusters, which are then independently flipped with probability $1/2$. A flipped spin is given by $\bm{S}' = \bm{S} - 2 S^{\parallel} \bm{S}_{\rm{ref}}$. An updated configuration is generated after all clusters have been considered.
The algorithm operates with a computational complexity of $\mathcal{O}(N)$, because each MC sweep involves evaluating $N(N-1)/2$ bonds, and on average, $\mathcal{O}(N)$ operations are required per spin.

In the 1990s, Luijten and Blöte introduced an efficient cluster algorithm for the long-range Ising model~\cite{Luijten1995}. Their method is based on the observation that, on average, only $\mathcal O(N)$ among $\mathcal O(N^2)$ bonds participate in the cluster construction. They proposed a rejection-free approach using the binary search to directly sample the next bond to activate (the bond activation event) instead of sequentially inspecting all bonds. This strategy reduces the number of operations per MC sweep to $\mathcal O(N \log N)$, resulting in a computational complexity of $\mathcal O(\log N)$, which drastically increases the efficiency of the algorithm. Nevertheless, their method is primarily limited to the LR Ising model. Additionally, the binary search requires a look-up table to sample the bond activation events, making it technically challenging to apply to 2D models, and special modifications to the coupling function are needed~\cite{Luijten2002}. 

To overcome these limitations, we integrate the clock Monte Carlo method with the LB algorithm. The clock Monte Carlo method replaces the conventional Metropolis acceptance criterion with a factorized Metropolis filter, which expresses the acceptance probability as a product of independent factors; an update is rejected if any individual factor fails. By directly sampling the first rejecting factor (first-rejection event) using the clock sampling technique, the computational complexity is significantly reduced from $\mathcal{O}(N)$ to $\mathcal{O}(1)$~\cite{michel2019}. Recognizing the similarity between first-rejection events and bond activation events, we apply the clock sampling technique to efficiently sample bond activations.

Following the Ref~\cite{michel2019}, we first introduce a \textit{configuration-independent} activation probability for each bond
\begin{equation}
\hat{\rho}_{ij} = 1 - e^{-2 \beta J_{ij}}.
\end{equation}
This probability depends solely on the strength of the interaction $J_{ij}$ and ensures that for any spin configuration, $\hat{\rho}_{ij} \geq \rho_{ij}$. Accordingly, $\hat{\rho}_{ij}$ is referred to as the \textit{bound activation probability} and activating a bond according to $\hat{\rho}_{ij}$ is called a \textit{bound activation}. The actual activation of a bond $(i,j)$ then proceeds in two steps: a bound activation is first attempted with probability $\hat{\rho}_{ij}$; if successful, one performs a resampling with the relative probability:
\begin{align}
\rho_{ij,\rm{rel}} = \rho_{ij} / \hat{\rho}_{ij}.
\label{eq:relative_prob}
\end{align}
Through this two-step process, an explicit evaluation of the bond condition Eq.~\eqref{eq:bond_evaluation} is only required when a bound activation occurs. We exploit this property to achieve efficient sampling of bond activations.

 {In a long-range $O(n)$ spin model, there are $(N-1)/2$ types of displacement vectors $\vec{r}$. For each $\vec{r}$, there are $N$ interactions $J(\vec{r})$ due to the translational invariance of the system with PBC, composing a total of $N(N-1)/2$ interaction pairs~\cite{blote1995}. All bonds corresponding to the same $\vec{r}$ then share an identical $\hat{\rho}$. To construct clusters, we first generate, for each type of $\vec{r}$, a list of bound-activated bonds.} Then, these bonds are resampled to determine the actual activation. The list can be efficiently generated by sampling the distribution $P(k) = \hat{\rho}(1-\hat{\rho})^{k-1}$, which is the probability that the $k$-th bonds are bound activated, while the preceding $k-1$ bonds are rejected and skipped. This method is particularly advantageous when the $\hat{\rho}$ is small, which is typically the case near $T_c$; most bonds are skipped before a bond is inspected for activation, significantly reducing the overall computational cost. For a type of bond that consists of $N$ bonds with $\hat{\rho}$, the bond activation process works as follows:

(a) Initially, set index $k = 0$.

(b) Calculate the index of the next bond to inspect using
\begin{equation} 
k \leftarrow k + 1 + \left \lfloor \frac{\ln(\texttt{rand})}{\ln(1-\hat{\rho})} \right \rfloor, 
\end{equation}
where \texttt{rand} is a uniform random number in $(0,1]$ and $\lfloor x \rfloor$ denotes the floor function that returns the integer part of $x$.

(c) If $k \le N$, attempt to activate the $k$-th bond with relative probability $\rho_{k,\text{rel}}$ in Eq,~\eqref{eq:relative_prob}.

(d) Repeat from step (b) until $k > N$.

The cluster construction process concludes once all types of bonds have been considered. This procedure leads to a considerably faster algorithm by reducing both the number of bond inspections and the number of random number generations. Compared to the LB algorithm, our method eliminates the need for a look-up table and alleviates truncation errors associated with discrete cumulative probability integration approximations. It faithfully reproduces the stochastic dynamics of the original multi-cluster method while achieving $\mathcal{O}(1)$ computational complexity.

In this work, we have performed extensive MC simulations of the LR XY model using the enhanced LB algorithm, with linear system sizes ranging from $L=16$ to $L=8192$. Each system is thermalized for $4\times10^3$ MC sweeps, which is sufficient to reach thermal equilibrium. For system sizes $L \le 4096$ at low temperatures $\beta=1,2,4,8$ and near the critical temperature of various $\sigma$, we collect over $2 \times 10^6$ samples to achieve high statistical accuracy. Under other parameters, where high-precise data is not required, the minimal number of samples is $4\times 10^5$. Simulations for the largest system size $L=8196$ are conducted at selected parameters, and the results are obtained from over $4\times10^4$ samples.  {Our simulation for the 2D LR XY model near $\sigma=2$ does not encounter strong critical slowing down issues.} Specifically, for a simulation with $L=8192$ and $\sigma = 2$ near the critical point $T_c$, the integrated autocorrelation time $\tau_{\text{int}}$ in the unit of MC sweep is $\sim 9$ for the energy-like quantity and $\sim 4.6$ for the squared magnetization. The average computational complexity is $\sim 2$ operations per spin, which remains virtually constant for all simulations near $T_c$, and it takes around $11$ CPU days to go over $10^4$ MC sweeps  {for a CPU with frequency $2.4$GHz.} Moreover, the CPU time per spin update is proportional to the computational complexity, and the space complexity of the algorithm scales linearly with system volume $N$. For the largest system size $L=8192$, our implementation requires $\sim 10.5$ gigabytes of memory.


\subsection{Sampled Quantities}
Various physical quantities are sampled in the simulation to investigate the properties of the model. After thermalization, the observables are measured after every MC sweep. For a given spin configuration, we sample the following observables:
\begin{enumerate}[(a)]
\item Total energy from the NN interactions, $\varepsilon = - J_{\text{nn}} \sum_{\langle i,j \rangle} \bm{S}_i \cdot \bm{S}_j$, where $\langle i,j \rangle$ denotes nearest neighboring pairs of spins and $J_{\text{nn}} = c(\sigma, N)J$ is the NN coupling strength.

\item Magnetization density $\mathbf{M} = L^{-2} \sum_{i=0}^N \bm{S}_i$, and its Fourier mode $\mathbf{M}_{k} = L^{-2} \left| \sum_i \bm{S}_i e^{i\bm{k} \cdot \bm{r}_i} \right|$, where $\bm{k}=\frac{2\pi}{L}\hat{\bm{x}}$, the smallest wave vector of the reciprocal lattice in $\bm{x}$-direction.

\end{enumerate}
The measurement consumes only a small fraction of the total simulation time.
We then obtain the ensemble average ($\langle \cdot \rangle$) of the following quantities:
\begin{enumerate} [(i)]
    \item The magnetic susceptibility $\chi=L^2\langle M^2\rangle$ and its Fourier mode $\chi_k=L^2\langle M_k^2\rangle$.
    
    \item The $2^{\text{nd}}$-moment correlation length~\cite{salas1998,caracciolo1993,tuan2022}, defined as \begin{align}\label{eq:2nd_moment_correlation}
        \xi_{\text{2nd}} = \frac{1}{2 \sin(|\bm{k}|/2)} \sqrt{\langle M^2 \rangle/\langle M_k^2 \rangle - 1}.
    \end{align}
 
    \item The Binder ratio of magnetization
    \begin{align}
        Q_{m} = \frac{\langle M^2\rangle^2}{\langle M^4\rangle}.
    \end{align}
    
    \item The scaled covariance between $\varepsilon$ and $M^2$, defined as
    \begin{align}
    K &=-\frac{L^2}{\langle M^2 \rangle}\left(\langle \varepsilon M^2 \rangle-\langle\varepsilon\rangle \langle M^2\rangle\right).
    \label{K_define}
    \end{align}

    \item The specific-heat-like quantity, defined as
    \begin{align}
        C_{\text{NN}} = \beta^2 L^2 \left( \langle \varepsilon^2\rangle - \langle \varepsilon  \rangle^2 \right).
    \end{align}
    where $\beta=1/T$ is the inverse temperature.
    
\end{enumerate}

Note that the second-moment correlation length $\xi_{\text{2nd}}$ is different from the commonly referred exponential correlation length $\xi_{\text{exp}}$. The former is defined based on the Fourier spectrum of the correlation function~\cite{caracciolo1993,edwards1989}, while the latter is obtained by fitting the exponentially decaying correlation function. Both correlation lengths are well-defined in the disordered phase and become asymptotically equivalent in the thermodynamic limit~\cite{Yao2025}. In practice, the second-moment correlation length $\xi_{\text{2nd}}$ can be easily measured in MC simulations, whereas determining $\xi_{\text{exp}}$ can be challenging  {because corrections simultaneously arising from both finite distance and finite system sizes and their mixing. In addition, near and at the critical point, $\xi_{\rm exp}$ becomes ill-defined since the correlation function exhibits a power-law scaling. In the long-range ordered phase, it is extremely challenging to extract $\xi_{\rm exp}$ since the correlation function saturates to a non-zero constant as the distance becomes infinitely large.} Throughout this work, unless specified otherwise, the correlation length $\xi$ refers to the second-moment correlation length $\xi_{\text{2nd}}$, which is well defined in all disordered, critical, and long-range ordered phases.

The dimensionless ratio $\xi/L$ is an effective tool for identifying the critical points and the phase transition properties~\cite{viet2009, ding2014, tuan2022}. In the disordered phase, the correlation length $\xi$ is finite, and the ratio $\xi/L$ vanishes as $L$ increases. At criticality, $\xi$ is proportional to the linear system size $L$ from the algebraic decaying behavior of the correlation function, and the ratio $\xi/L$ takes a universal value in the thermodynamic limit. In the ordered phase, $\xi/L$ diverges since $M^2$ remains finite while $M_k^2$ vanishes with increasing $L$. As a consequence, for a continuous phase transition separating a long-range ordered from a disordered phase, the intersection of $\xi/L$ curves for different $L$ accurately pinpoints the critical point. In the case of a BKT transition, however, the ratio converges to a smooth and universal curve since the whole low-T QLRO phase is essentially critical with an algebraically decaying correlation function.

Throughout the simulation, the error bars are estimated using standard error analysis techniques: for simple observables, such as $\varepsilon, M^2$, and $M_k^2$, the error bars are calculated using the standard binning method; for composite observables obtained by the arithmetic combinations of simple observables, such as the 2nd-moment correlation length $\xi$ and Binder ratio $Q$, the error bars are estimated using the standard jackknife method.

\subsection{Finite-size Scaling Analysis}
\label{FSS Analysis}
Finite-size scaling (FSS) analysis is an essential technique in numerical investigation of critical phenomena. In the language of the renormalization group (RG) theory, the critical behavior of systems belonging to the same universality class is governed by a common fixed point. Near critical points, the free energy, and thus other physical observables, can be expressed as universal functions of scaling fields. FSS extends these universal functions to finite systems, enabling precise determination of critical points and accurate extraction of critical exponents.
Near the critical point, the free-energy density $f(t,h,u, L^{-1})$ is a function of the thermal scaling field $t$, the magnetic scaling field $h$, an irrelevant field $u$, and the finite-size field $L^{-1}$. Then, the FSS behavior of free-energy density is given by,
\begin{equation}
    f(t,h,u,L^{-1})=L^{-d}f(tL^{y_t},hL^{y_h},uL^{y_i},1) + g(t,h),
    \label{eq:free_energy}
\end{equation}
where $y_t=1/\nu$ is the thermal renormalization exponent ($\nu$ is the correlation-length exponent); $y_h$ is the magnetic renormalization exponent; $y_i<0$ is the leading irrelevant scaling exponent; and $g$ is the analytic part of free-energy density. The scaling fields are non-universal functions of microscopic interactions and external physical parameters. To the lowest order, $t$ is proportional to $T-T_c$, i.e., the distance between temperature and critical temperature; $h$ is proportional to the physical external magnetic field $H$. The irrelevant field $u$ reflects the distance of criticality of the system and the fixed point.

The FSS behavior of various physical observables near $T_c$ can be derived from Eq.~\eqref{eq:free_energy}. Here, we present the scaling form of several key quantities: (a) magnetic susceptibility $\chi$, (b) scaled covariance $K$, and (c) dimensionless quantities, such as $Q_{m}$ and $\xi/L$.

\textbf{(a) FSS of $\chi$.}
The magnetic susceptibility $\chi$ is calculated from the averaged squared magnetization $\langle M^2 \rangle$, which is propositional to $\partial^2 f / \partial h^2$ at $h=0$. Thus, near the critical point, the magnetic susceptibility scales as
\begin{align}
    \chi = L^{2y_h - d} \tilde{\chi}(L^{y_t} t ,L^{y_i} u, 1) + \cdots,
    \label{chi_scale2}
\end{align}
where $\tilde{\chi}$ is a universal function, and the contribution due to $g$ is not specified. The leading exponent $2y_h-d$ can be replaced by $2-\eta$ where $\eta$ is the anomalous dimension of the correlation function at the critical point.

\textbf{(b) FSS of $K$.}
The quantity $K$ is proportional to the correlation between the squared magnetization $M^2$ and the NN energy $\varepsilon$.
The NN energy $\varepsilon$ shares a similar scaling behavior as the total energy $E$ near $T_c$ as $\sim L^{y_t-d}$~\cite{blote1995,deng2003}. Consequently, the correlation between $M^2$ and $\varepsilon$ is proportional to the temperature derivative of squared magnetization $\frac{\partial}{\partial t}\langle M^2\rangle$~\cite{blote1995}.
The FSS form of $K$ at the critical point $t=0$ is then given by
\begin{align}
    K&\propto\frac{1}{\langle M^2\rangle}\frac{\partial}{\partial t}\langle M^2\rangle\Big|_{t=0} = \frac{1}{\chi}\frac{\partial}{\partial t}\chi\Big|_{t=0}  \\
    &\approx a_0 L^{y_t}(1+a_1L^{y_u}) + a_2 L^{y_t+d-2y_h} +\cdots,
\label{eq:K_scale}
\end{align}
where $a_i$ are non-universal constants. The leading $L^{y_t}$ scaling behavior in Eq.~\eqref{eq:K_scale} makes $K$ an excellent quantity to estimate the thermal scaling exponent $y_t=1/\nu$.

\textbf{(c) FSS of dimensionless ratios.} 
Dimensionless ratios, such as Binder ratio $Q_{m}$ and correlation length ratio $\xi/L$, have been widely used in FSS analysis to locate the critical point of phase transitions~\cite{viet2009, komura2012, ding2014}. Near the critical point, the Binder ratio follows the scaling behavior
\begin{equation}
    Q_m(t,u,L)=Q_m(tL^{y_t},uL^{y_u},1)+\cdots,
    \label{eq:Qm_scale}
\end{equation}
where the ellipsis denotes omitted correction terms arising from the field dependence of the analytic part of the free energy and higher-order contributions.
Similary, since the correlation length $\xi$ scales as the system size $L$ near $T_c$, the FSS form of correlation length ratio $\xi/L$ is given by,
\begin{align} \label{eq:corrL_scale}
    \xi/L \sim \tilde{\xi}(tL^{y_t},uL^{y_u},1),
\end{align}
where $\tilde{\xi}(\cdot)$ is a universal scaling function.

\section{Overview of Results}\label{sec:results_phase_diagram}


\begin{figure*}[t]
    \centering
    \includegraphics[width=\linewidth]{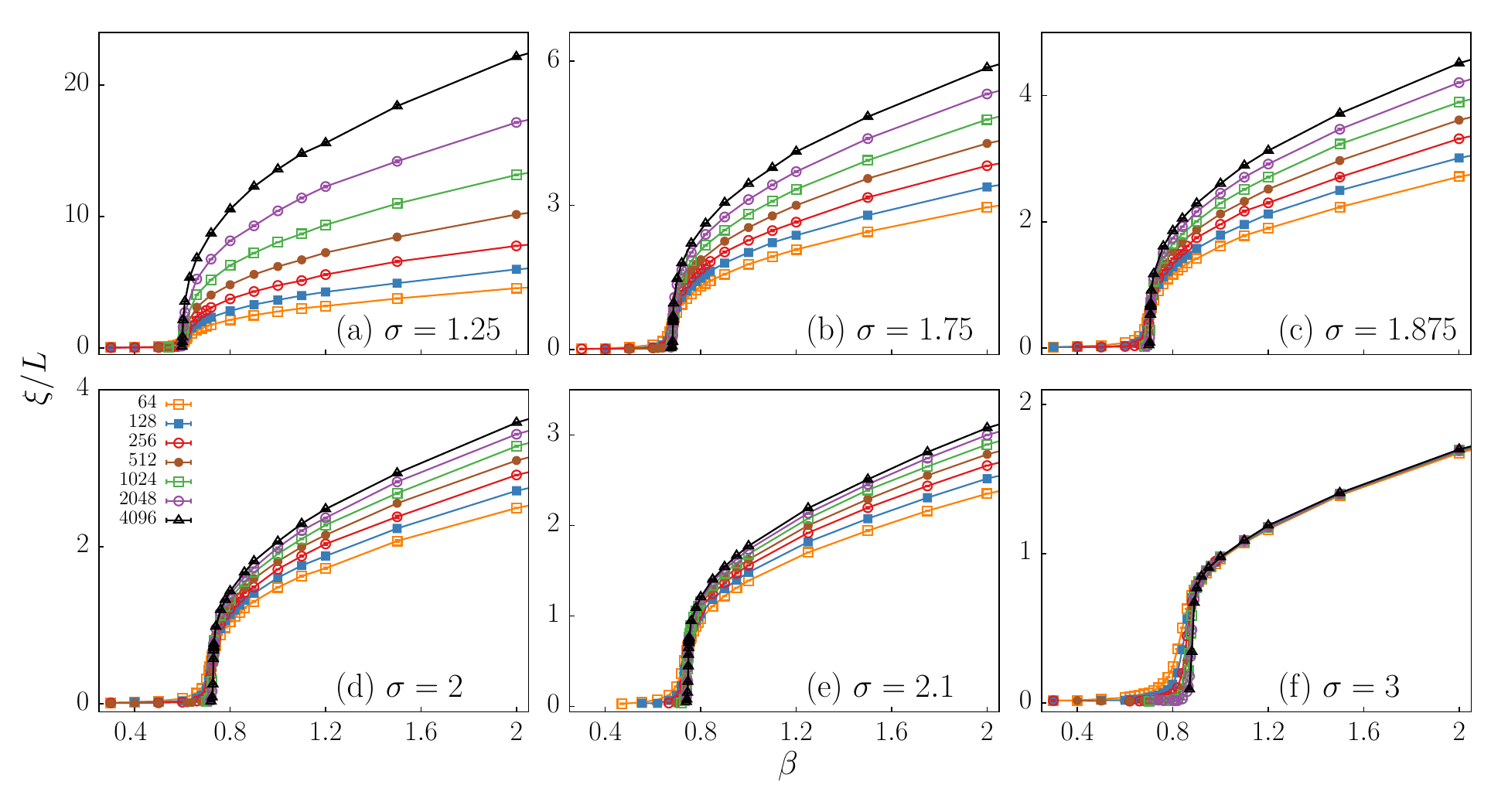}
    \caption{Overview of the dimensionless ratio $\xi/L$ for different $\sigma$ values. The phase transition type changes from a second-order transition for $\sigma \le2$ to a BKT-type transition for $\sigma > 2$, as shown here. The second-moment correlation length divided by $L$, $\xi/L$, is plotted as a function of inverse temperature $\beta$ for various $\sigma$ values: (a) $\sigma = 1.25$, (b) $\sigma = 1.75$, (c) $\sigma = 1.875$, (d) $\sigma = 2$, (e) $\sigma = 2.1$, and (f) $\sigma = 3$. System sizes range from $L = 64$ to $L = 4096$. For $\sigma \le 2$, the curves for different system sizes exhibits a clear intersection, indicating a second-order phase transition. For $\sigma = 3$, all $\xi/L$ curves for $L > 64$ converge after the transition point, a hallmark of a BKT transition. In the $\sigma = 2.1$ case, smaller system sizes (up to $L = 1024$) show a crossing due to the strong finite-size effect. For larger system sizes ($L = 2048$ and $L = 4096$), the $\xi/L$ values tend to converge in the low-T phase. This subtle case will be elucidated in Sec.~\ref{sigma_g2}.
    }
    \label{Q_crossing}
\end{figure*}

Before delving into the detailed analysis of the LR XY model, we provide an overview of the main findings to facilitate a better understanding. As the long-range model enters the nonclassical regime in $\sigma \le\sigma_*$, the critical behavior differs from that of the short-range regime. For the LR Ising model, the universal properties of the transition, such as the critical exponents and other universal quantities, vary as $\sigma$ in the nonclassical regime. Therefore, previous studies of the LR Ising model focus on the accurate extraction of critical exponents or values of universal ratios like Binder cumulant, which, however, can be extremely subtle and challenging~\cite{Luijten2002,picco2012,horita2017}. In contrast, for the LR XY model, we expect a much more evident signature when the system enters the nonclassical regime --- the type of transition changes from a BKT transition to a second-order transition. The change in transition type leads to qualitatively distinct behaviors in both the low- and high-T phases, as well as in the critical region. As a result, determining the crossover point $\sigma_*$ becomes considerably easier in the LR XY model than in the LR Ising case.

Based on extensive MC simulations, we identify three distinct regimes in the phase diagram of the LR XY model: the \textit{classical} regime ($\sigma \leq 1$), the \textit{nonclassical} regime ($1 < \sigma \leq 2$), and the \textit{short-range} regime ($\sigma > 2$), as illustrated in Fig.\ref{PD}. For $\sigma < 1$, the system undergoes a second-order transition to a LRO low-T phase, with critical behavior governed by Gaussian mean-field theory~\cite{defenu2023}. In the short-range regime with $\sigma > 2$, the system exhibits a standard BKT transition to a QLRO phase. The nonclassical regime ($1 < \sigma \leq 2$) is particularly noteworthy, where detailed FSS analysis reveals a second-order phase transition to a ferromagnetic phase at low T with $\sigma$-dependent critical exponents depending. This finding contrasts with the previously suggested phase diagram based on field theoretical analysis --- the proposed intermediate QLRO phase in the regime $1.75 < \sigma \le 2$ is not observed from our data~\cite{Giachetti2021, Giachetti2022, note}. Moreover, our results suggest the crossover point between SR and LR universality is at $\sigma = 2$, rather than at $\sigma = 1.75$ as predicted by Sak's criterion.

The second-order phase transition in the nonclassical regime is strongly hinted in Fig.~\ref{Q_crossing} where the correlation length ratio $\xi/L$ is plotted as a function of the inverse temperature $\beta$. Figure~\ref{Q_crossing} presents $\xi/L$ versus $\beta$ for $\sigma=1.25$, $1.75$, $1.875$, $2$, $2.1$ and $3$. The data from system sizes ranging from $L=64$ to $L=4096$ are shown for a comprehensive examination of finite-size effects. For $\sigma \le 2$, $\xi/L$ curves of different $L$s clearly exhibit a universal intersection at some finite temperature, and in the low-T phase $\xi/L$ diverges as the system size increases, reflecting the emergence of long-range order. This typical scaling behavior indicates a second-order phase transition into an LRO phase, and the crossing points mark the location of the critical points $T_c$. This differs from the BKT transition in the SR regime, for which the results are also shown as a comparison. For $\sigma = 3$, the $\xi/L$ curves do not show a crossing; instead, curves for system sizes greater than $L = 64$ converge to a universal function at low temperatures, confirming a BKT transition to a QLRO phase. However, at $\sigma = 2.1$, the behavior is more nuanced. While the $\xi/L$ curves of smaller system sizes (up to $L=1024$) seemly suggest a crossing, those for larger sizes ($L=2048$ and $L=4096$) start to converge after the critical point. Such behavior reflected the strong finite-size effect near the boundary between LR and SR universality.  {A systematic and detailed analysis will be given later. We will show that, in the low-T phase, the squared correlation-length ratio, $(\xi/L)^2$, diverges as $\sim L^{2-\sigma}$ for $\sigma <2$ and logarithmically as $\sim \ln L$ for $\sigma=2$; in contrast, $\xi/L$ statures to some finite and $T$-dependent value for $\sigma=2.1$. In the high-T phase when approaching the critical point, the correlation length $\xi$ exhibits power-law divergence for $\sigma\leq2$, indicative of a second-order phase transition; in contrast, $\xi$ diverges exponentially for $\sigma=2.1$ and $3$, indicative of a BKT phase transition. Moreover, the fitted critical exponents also separate $\sigma\leq2$ from the SR regime.}

\section{Long-range order and Goldstone mode at low temperature}
\label{sec:results_I}

In this section, we conduct a comprehensive study of the system's low-T properties. In subsection~\ref{LRO_s2}, we show the existence of LRO in the low-T phase for $\sigma \leq 2$. 
In subsection~\ref{Goldstone mode}, we investigate the Goldstone mode excitations in the low-T phase for $\sigma < 2$, demonstrating that the correlation function adheres to the form $g(r)=g_0+cr^{-\eta_\ell}$ where $\eta_\ell=2-\sigma$. For $\sigma = 2$, we propose a possible logarithmic behavior and verify it numerically. Subsequently, the QLRO phase at low temperatures for $\sigma>2$ is discussed in subsection~\ref{sigma_g2}. Finally, we revisit the boundary case of $\sigma=2$ and examine its distinctive finite-size scaling in subsection~\ref{sigma_e2}. Our research primarily focuses on $\sigma$ values of $1.75,1.875,2$ at various low temperatures characterized by $\beta=1,2,4,8$. Additionally, cases of $\sigma=2.1,3$ are partially included for comparison. 

\subsection{Long-Range Order for $\sigma\leq 2$}
\label{LRO_s2}

\begin{figure}[t]
    \centering
    \includegraphics[width=\linewidth]{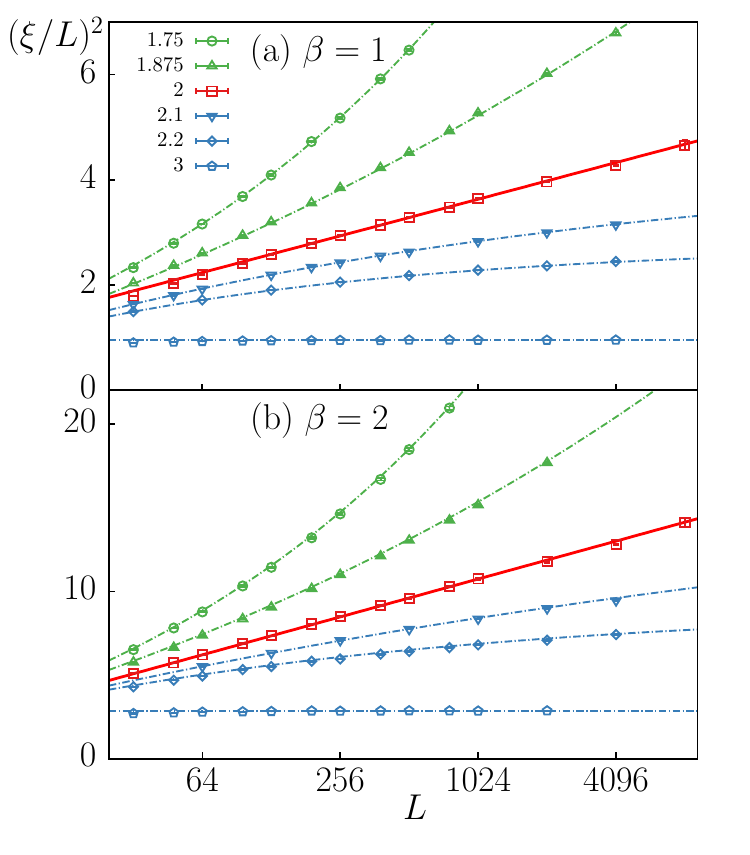}
    \caption{Different behaviors of $(\xi/L)^2$ in the low-T phase for various $\sigma$. $(\xi/L)^2$ as a function of system size $L$ in the semi-logarithmic coordinates is shown for various $\sigma$ values at inverse temperatures $\beta = 1$ (a) and $\beta = 2$ (b).  {The linear behavior of the red squares demonstrates the logarithmic divergence of $(\xi/L)^2$ for the case of $\sigma = 2$. For $\sigma<2$ (the green dots in the figure), $(\xi/L)^2$ diverges faster, indicating a possible power-law divergence. This is confirmed in Sec.~\ref{Goldstone mode}. The diverging $(\xi/L)^2$ is a signature of the LRO in the low-T phase for $\sigma\leq2$ . For $\sigma > 2$ (the blue dots), $(\xi/L)^2$ tends to converge, indicating QLRO in the system.}}
    \label{correL_lowT_s}
\end{figure}

In this part, we demonstrate the existence of LRO in the low-T phase for $\sigma \le 2$. We first study the scaling behavior of the squared correlation length ratio $(\xi/L)^2$, which exhibits distinct behavior in the LRO and QLRO phases. Specifically, for given $T<T_c$, as the system size increases, this ratio diverges in the LRO phase, while in a QLRO phase, it converges to some non-zero constant. Fig.~\ref{correL_lowT_s} displays $(\xi/L)^2$ as a function of $L$ at inverse temperatures $\beta = 1$ and $\beta = 2$ in a semi-logarithmic scale, revealing the distinct low-T scaling behaviors of $(\xi/L)^2$ for various values of $\sigma$. For $\sigma < 2$, the curves bend upwards as $L$ increases, and this divergence can be well-fitted by a power-law function, which will be discussed later. For $\sigma = 2$, the data points fall onto a straight line, suggesting that $(\xi/L)^2$ diverges logarithmically with $L$. For both cases, despite the different forms of scaling, $(\xi/L)^2$ diverges as $L$ increases, indicating the presence of LRO in the low-T phase for $\sigma \le 2$. In contrast, for $\sigma>2$, the system is in the QLRO phase at low temperatures, and $(\xi/L)^2$ is expected to converge. This can be seen clearly in the case of $\sigma=3$ at both $\beta=1$ and $2$ where $(\xi/L)^2$ rapidly converges to a constant as $L$ increases. However, for $\sigma = 2.1$ and $2.2$, the proximity to the marginal case $\sigma = 2$ results in strong finite-size effects, leading to a slow convergence of $(\xi/L)^2$. As shown in Fig~\ref{correL_lowT_s}, for both $\beta=1$ and $2$, $(\xi/L)^2$ curves slowly increase at a slackening rate, and we expect them to finally converge to a constant. In subsection~\ref{sigma_g2}, we further verify this expectation. 

\begin{figure*}[t]
    \centering
    \includegraphics[width=\linewidth]{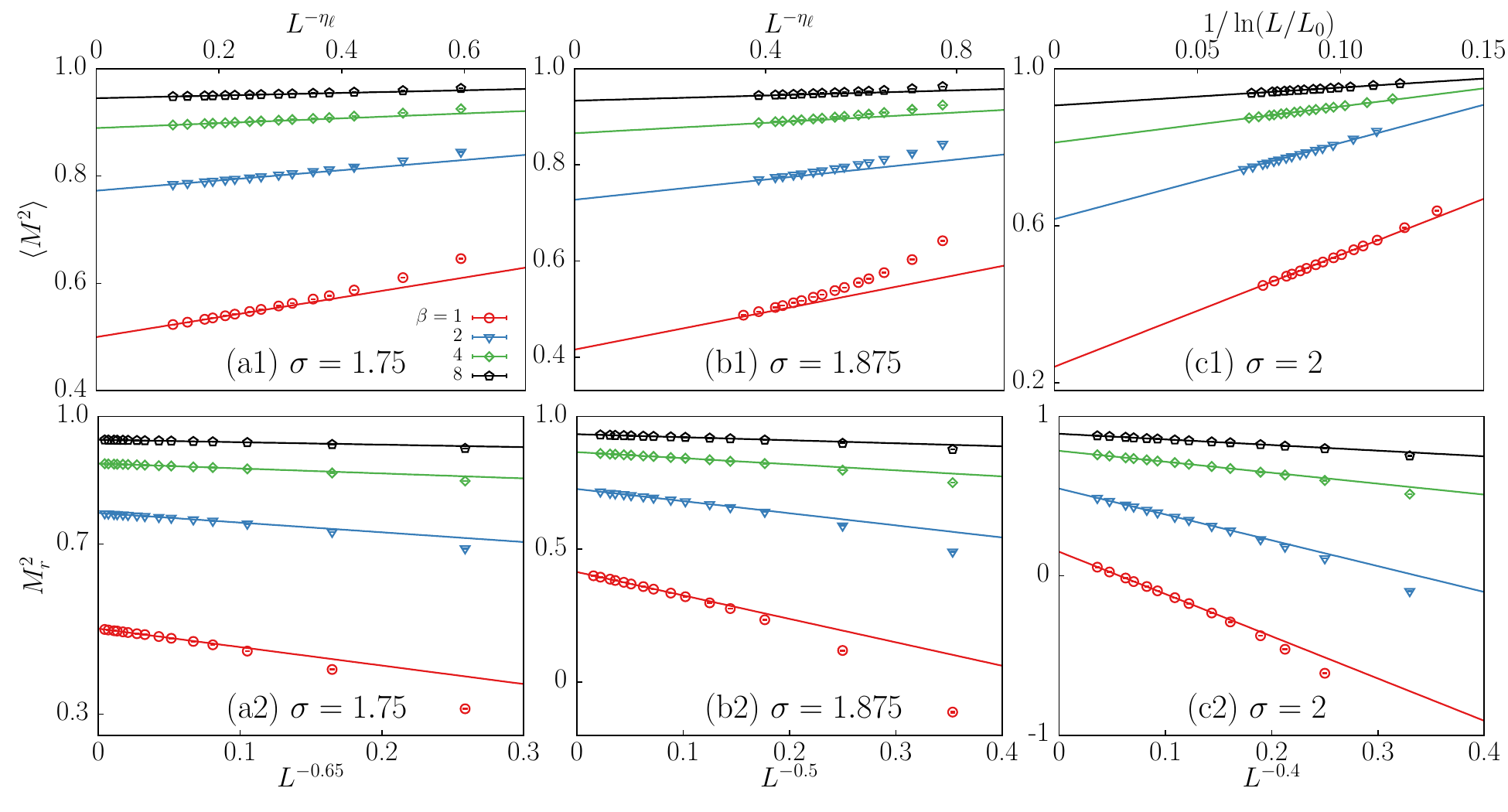}
    \caption{The existence of spontaneous magnetization for $\sigma \leq 2$ in low-T phase. The squared magnetization $\langle M^2\rangle$ is plotted as a function of system size $L$ for various temperatures at $\sigma = 1.75$ (a1), $\sigma = 1.875$ (b1), and $\sigma = 2$ (c1). The $x$-axis represents the leading finite-size correction term, which varies by panel: $L^{-\eta_\ell}$ ($\eta_\ell = 0.25, 0.125$ for $\sigma = 1.75, 1.875$, respectively) in panels (a1) and (b1), $1 / \ln(L / L_0)$ in panel (c1). The fitting parameter $L_0=e^{-5.4}, e^{-6.8}, e^{-6.38}, e^{-6.2}$ for $\beta = 1, 2, 4, 8$ in panel (c1). For all three panels, the fitting lines align with the data point, and intersect with y-axis at positive values, indicating the spontaneous magnetization at low temperatures in the thermodynamic limit.
    To diminish the finite-size correction, the plots of $M_r^2 = \langle M^2\rangle-b\langle M_k^2\rangle$ are presented at $\sigma=1.75$ (a2), $\sigma=1.875$ (b2) and $\sigma=2$ (c2). The $x$-axis represents the leading correction term: $L^{-0.65}$ in panel (a2), $L^{-0.5}$ in panel (b2), and $L^{-0.4}$ in panel (c2). $b$ equals $22$ for panel (a2), 48 for panel (b2), $149,175,152,154$ in $\beta=1,2,4,8$ respectively for panel (c2). After subtracting a positive quantity, $M_r^2$ converges faster and continues to reach a positive value at the thermodynamic limit with smaller finite-size corrections, which is strong evidence of spontaneous magnetization.}
    \label{M2_Mr2}
\end{figure*}

 {The Monte Carlo data of the squared magnetization $\langle M^2\rangle$ are presented in Fig.~\ref{M2_Mr2} (a1-c1) for various $\beta$s. The squared magnetization is plotted against its leading correction term (will be explained below), demonstrating the extrapolation of $\langle M^2\rangle$ to the $L\rightarrow \infty$ limit. For $\sigma=1.75,1.875$ and $2$, at $\beta\geq1$, $\langle M^2\rangle$ tends to converge to nonzero values in the thermodynamic limit, which strongly indicates the existence of LRO in the low-T phase for $\sigma \le2$. It will be shown that $\langle M^2\rangle$ follows a power-law convergence for $\sigma<2$ and a logarithmic convergence for $\sigma=2$, and the corresponding fitting form of $\langle M^2\rangle$ is explained below.}

For the LR XY model, due to the spontaneous breaking of continuous symmetry, the LRO phase shall exhibit Goldstone modes. It is shown in Sec~\ref{Goldstone mode} that the correlation function has the form: $g(x) = g_0 +c\cdot x^{-\eta_\ell}$ where $\eta_\ell=2-\sigma$. Therefore, considering the relationship between $\langle M^2\rangle$ and correlation function: $\langle M^2\rangle\sim \frac{1}{L^2}\int g(r)\mathrm{d}^2r$, $\langle M^2 \rangle$ is fitted with the following ansatz,
\begin{equation}
    \langle M^2\rangle = g_0+L^{-\eta_\ell}(a_0+a_1L^{-\omega}),
\label{M2_scaling}
\end{equation}
where $a_1L^{-\omega}$ term accounts for additional corrections. 
For $\sigma=2$, however, the correlation function can't be easily obtained from a Goldstone mode analysis due to the degeneracy between the LR and the SR interaction terms; nevertheless, in subsection~\ref{Goldstone mode}, we propose a possible scaling form at $\sigma=2$: $\chi_k\sim L^2\ln(L/L_0)$ or $\langle M_k^2\rangle\sim \ln(L/L_0)$. Since $M_k$ is the Fourier mode of $M$, $\langle M^2\rangle$ is expected to exhibit similar scaling behavior, with an additional constant term. Thus for $\sigma=2$, $\langle M^2\rangle$ is fitted as
\begin{equation}
    \langle M^2 \rangle = g_0 + \frac{a_0}{\ln(L/L_0)} .
\label{M2_scaling_log}
\end{equation}
A detailed discussion of such logarithmic behavior in the marginal case of $\sigma = 2$ is provided in the next subsection~\ref{Goldstone mode}.

 {In Fig.~\ref{M2_Mr2} (a1) and (b1) ($\sigma=1.75$ and $1.875$), we plot the $x$-axis as $L^{-\eta_\ell}$, the leading correction term according to Eq.~\eqref{M2_scaling}. The plots reveal an asymptotic linear relationship, except for the small system size data, which is due to finite-size effects. The non-zero intercepts in the $y$-axis reveal the presence of spontaneous magnetization in the low-T phases. Fig.~\ref{M2_Mr2} (c1) illustrates the logarithmic scaling $\langle M^2 \rangle$ for $\sigma = 2$, where we plot $\langle M^2\rangle$ versus $1 / \ln(L / L_0)$, with $L_0$ a fitting parameter. The apparent linear relationship confirms the logarithmic scaling of $\langle M^2\rangle$ and the data fit well to the ansatz in Eq.~\eqref{M2_scaling_log}. Again, the spontaneous magnetization is evident from the non-zero intercept.} 

Extrapolating $\langle M^2 \rangle$ to the thermodynamic limit in the first three panels in Fig.~\ref{M2_Mr2} demonstrates the existence of LRO in the low-T phase for $\sigma \leq 2$. However, due to the large finite-size effect, for $\sigma=1.875$ and $2$, $\langle M^2\rangle$ converges slowly. To mitigate the finite-size corrections and better observe the LRO, we define the residual squared magnetization,
\begin{equation}
M_r^2 = \langle M^2 \rangle - b \cdot \langle M_k^2 \rangle,
\end{equation}
where $b>0$ is some constant to be determined. Here, $\langle M_k^2\rangle$ is the squared Fourier mode of magnetization, which is closely related to the Fourier transform of the correlation function, $\langle M_k^2\rangle\sim \frac{1}{L^2}\int g(r)e^{i\boldsymbol{k}\boldsymbol{r}}\mathrm{d}^2r$. Therefore, for $\sigma<2$, $\langle M_k^2\rangle$ scales as 
\begin{equation}
    \langle M_k^2\rangle = L^{-\eta_\ell}(a_0+a_1L^{-\omega}),
\label{Mk2_scaling}
\end{equation}
where $a_1L^{-\omega}$ attributes to correction terms. A notable difference between the scaling behavior of $\langle M^2 \rangle$ and $\langle M_k^2 \rangle$ is that $\langle M_k^2 \rangle$ vanishes in the thermodynamic limit. Similarly for $\sigma = 2$, we expect $\langle M_k^2 \rangle$ behaves as,
\begin{equation}
    \langle M_k^2 \rangle = \frac{a_0}{\ln(L/L_0)}.
\label{Mk2_scaling_log}
\end{equation}
A detailed discussion on this scaling behavior will be provided in the next subsection \ref{Goldstone mode}. Since $\langle M_k^2 \rangle$ shares the same scaling as the leading correction term in $\langle M^2 \rangle$, selecting an appropriate value for $b$ can effectively cancel out the leading correction, thus reducing the finite-size effects in $M_r^2$.  {More importantly, since $b$ is a positive constant, $M^2_r$ act as a lower bound of $\langle M^2\rangle$, and in the $L \rightarrow \infty$ limit, one has $M_r^2 = \langle M^2\rangle$. Hence, if $M_r^2$ converges to a positive value in the thermodynamic limit, $\langle M^2\rangle$ must also be positive.}

The lower three panels of Fig.~\ref{M2_Mr2} display $M_r^2$ as a function of $L^{-\omega}$ for $\sigma = 1.75, 1.875,2$, with the values of $b$ and $\omega$ provided in the figure caption. With reduced finite-size corrections, the data points converge quickly to the extrapolated thermodynamic-limit value. The extrapolated curve clearly intersects the $y$-axis at positive values, indicating finite magnetization density. The discrepancies at smaller $L$ values indicate the presence of additional, smaller corrections beyond $L^{-\omega}$. Furthermore, the coefficient of the leading correction term of $M_r^2$ is negative, causing $M_r^2$ to increase with the size of the system. In all cases, the values of $M_r^2$ for the largest system size are already positive. Consequently, in the $L \rightarrow \infty$ limit, one must have $\langle M^2\rangle \geq M_r^2 > 0$. Thus, we conclude that the system retains long-range order in the low-T phase for $\sigma \leq 2$. 

The fitting detail of $M_r^2$ is shown below. Firstly, the fitting of Eq.~\eqref{M2_scaling} and Eq.~\eqref{Mk2_scaling_log}, provides the the amplitude of the correction term we aim to eliminate, namely, $a_0$. The ratio of $a_0$ in $\langle M^2 \rangle$ and $\langle M_k^2 \rangle$ determines the value of $b$ in $M_r^2$. $M_r^2$ is then fitted with the ansatz,
\begin{equation}
M_r^2 = g_0 + a_1 L^{-\omega},
\end{equation}
where $g_0$ is the squared magnetization in the thermodynamic limit, and $\omega$ is the subleading correction exponent, which remains relatively constant across different temperatures. 

It is worth noting that our conclusion partially contradicts the theoretical arguments presented in Ref.~\cite{PhysRevLett.87.137203}, stating that the 2D LR XY and Heisenberg system cannot sustain LRO at $\sigma=2$. This discrepancy between the theoretical predictions and our numerical evidence underscores the marginal case at $\sigma=2$ as an open and intriguing problem.

\subsection{Goldstone Mode}
\label{Goldstone mode}

\begin{figure*}[th]
    \centering
    \includegraphics[width=\linewidth]{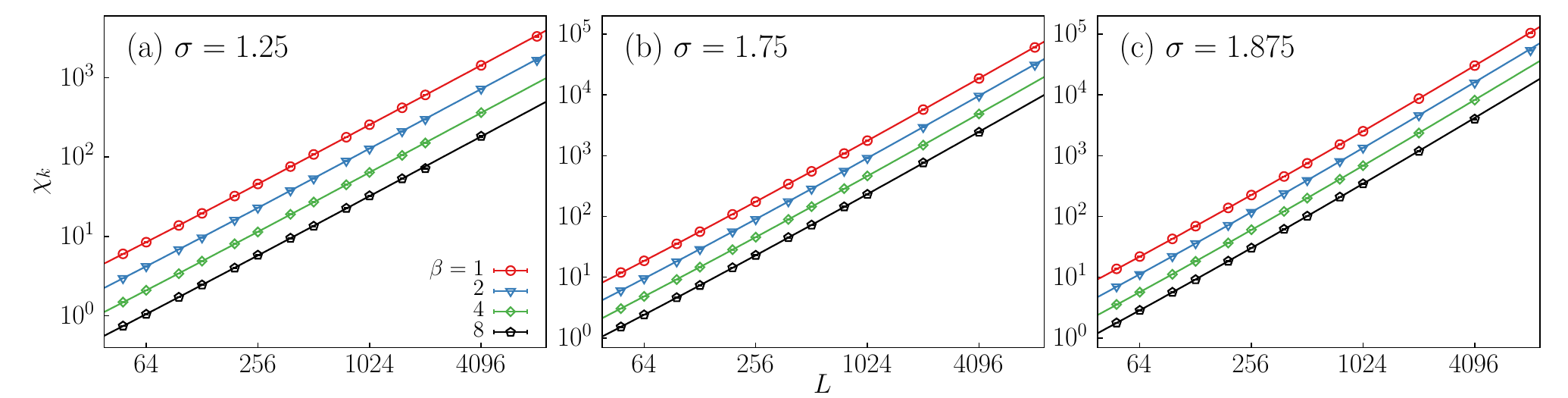}
    \caption{The demonstration of the presence of Goldstone mode. The log-log plots of $\chi_k$ versus system size $L$ are shown for different temperatures at $\sigma = 1.25$ (a), $1.75$ (b), and $1.875$ (c). As $L$ increases, the data points approach power-law growth with increasing system size $L$, confirming the form of the correlation function Eq.~\eqref{correlation function} and the presence of the Goldstone mode. Furthermore, the fitting lines at different temperatures are parallel for a certain $\sigma$, aligning with the prediction that the value of $\eta_\ell$ remains independent of temperature.}   
    \label{chi_k}
\end{figure*}

As discussed in the previous subsection, the spontaneous breaking of continuous symmetry leads to the emergence of Goldstone mode excitations in the LRO phase~\cite{Kardar}. In this subsection, we provide both theoretical arguments and numerical evidence for the algebraic decay of the correlation function in the low-T phase for $\sigma<2$, i.e., $g(x)=g_0+c\cdot x^{-\eta_\ell}$ with $\eta_\ell=2-\sigma$. Here, $x$ and $g(x)$ represent the distance and correlation between two sites. This power-law behavior arises from the low-energy Goldstone mode excitations. We also examine the marginal case $\sigma = 2$, where we conjecture a logarithmic decay of correlations and support this hypothesis through numerical analysis.

\textit{Theoretical derivation}.
The reduced Hamiltonian of long-range O$(n)$ spin model can be written in momentum-space as \cite{Fisher1972}:
\begin{equation}
\begin{aligned}
    \beta H=&\int{\frac{\mathrm d^dq}{(2\pi)^d}(\frac t2+\frac {K_2}{2} q^2+K_\sigma q^\sigma)\boldsymbol{\Psi}(\boldsymbol{q})\cdot\boldsymbol{\Psi}(-\boldsymbol{q})}+\\
    &\int\frac{\mathrm{d}^dq_1}{(2\pi)^d}\int\frac{\mathrm{d}^dq_2}{(2\pi)^d}\int\frac{\mathrm{d}^dq_3}  {(2\pi)^d}u \\
    & \boldsymbol{\Psi}(\boldsymbol{q_1})\cdot\boldsymbol{\Psi}(\boldsymbol{q_2})\cdot\boldsymbol{\Psi}(\boldsymbol{q_3})\cdot\boldsymbol{\Psi}(-\boldsymbol{q_1}-\boldsymbol{q_2}-\boldsymbol{q_3})
\end{aligned}
\label{Hamiltonian}
\end{equation}
where $\boldsymbol{q}$ denotes a $d$-dimensional momentum variable and $\boldsymbol{\Psi}(\boldsymbol{q})$ is the Fourier transform of a locally defined $n$-component spin field $\boldsymbol{\Psi}(\boldsymbol{x})$ ($\boldsymbol{\Psi}(\boldsymbol{x})$ and $\boldsymbol{\Psi}(\boldsymbol{q})$ 
respectively represent the spin field in real space and momentum space. For simplicity, they share the same symbol, and we hope this won't lead to any confusion); $t, K_2, K_\sigma$, and $u$ are interacting parameters, and $t$ varies linearly with the distance to criticality. 

In the last subsection, we have demonstrated the existence of spontaneous magnetization and continuous symmetry breaking at low temperatures for $\sigma\leq2$. Thus, we can adapt the mean-field approximation and consider small transverse fluctuations around it:
\begin{equation}
    \boldsymbol{\Psi}(\boldsymbol{x})=\overline{\Psi}\boldsymbol{\hat{e}_l}+\Psi_T(\boldsymbol{x})\boldsymbol{\hat{e}_t}
    \label{spinField}
\end{equation}
where $\boldsymbol{\hat e_l}$ represents the longitudinal direction vector along the mean-field spin direction, while $\boldsymbol{\hat e_t}$ denotes the transverse direction vector; $\overline{\Psi}=\sqrt{\frac{-t}{4u}}$ is the result of the mean-field approximation and $\Psi_T$ refers to the transverse fluctuation of magnetization in the transverse directions.
Up to the second order, the reduced Hamiltonian becomes~\cite{Kardar}:
\begin{equation}
    \beta H=V(\frac t2{\overline{\Psi}}^2+u{\overline{\Psi}}^4)+\frac1V\sum_{\boldsymbol{q}}(K_\sigma q^\sigma+\frac{K_2}{2}q^2)\cdot|\Psi_T(\boldsymbol{q})|^2
\label{betaH}
\end{equation}
where $V$ denotes the volume of the system. For $\sigma<2$, $K_\sigma q^{\sigma}$ is the leading term, and hence $\frac{K_2}{2} q^2$ can be neglected. Thus, the probability of a particular fluctuation configuration has the Gaussian form:
\begin{equation}
    P(\{\Psi_T(\boldsymbol{q})\})\propto e^{-\beta H}\propto \prod_{\boldsymbol{q}}\exp{(\frac {K_\sigma}{V}q^\sigma\cdot|\Psi_T(\boldsymbol{q})|^2)}.
\end{equation}
Hence, the two-point correlation function in the momentum space is:
\begin{equation}
\langle\Psi_T(\boldsymbol{q})\Psi_T(\boldsymbol{q'})\rangle=\frac{\delta_{\boldsymbol{q},-\boldsymbol{q'}}V}{2K_\sigma q^\sigma},
\end{equation}
and in the real space is:
\begin{equation}
\begin{aligned}
    \langle\Psi_T(\boldsymbol{x})\Psi_T(\boldsymbol{x'})\rangle&=\frac1{V^2}\sum_{\boldsymbol{q}, \boldsymbol{q'}}\langle\Psi_T(\boldsymbol{q})\Psi_T(\boldsymbol{q'})\rangle e^{i\boldsymbol{q}\cdot\boldsymbol{x}+i\boldsymbol{q'}\cdot\boldsymbol{x'}}\\
    &=\frac1V\sum_{\boldsymbol{q}}\frac{e^{i\boldsymbol{q}(\boldsymbol{x}-\boldsymbol{x'})}}{2K_\sigma q^\sigma}\\
    &=\frac1{2K_{\sigma}}\int\frac{\mathrm{d}^dq}{(2\pi)^d}\frac{e^{i\boldsymbol{q}(\boldsymbol{x}-\boldsymbol{x'})}}{q^\sigma}.
\end{aligned}
\end{equation}
Considering the integration $ \int\frac{\mathrm{d}^dq}{(2\pi)^d}\frac{e^{i\boldsymbol{q}\cdot\boldsymbol{x}}}{q^\sigma}$, divide the integral into angular and radial components and we can analyze its scaling to $x$ without explicitly computing it (Note that our derivation is under the condition of $\sigma<2=d$):
\begin{equation}
\begin{aligned}
    \int\frac{\mathrm{d}^dq}{(2\pi)^d}\frac{e^{i\boldsymbol{q}\cdot\boldsymbol{x}}}{q^\sigma}&=\int\frac{\mathrm d\Omega}{(2\pi)^d}\int\mathrm dq\frac{e^{iqx\cos\theta}}{q^{\sigma-d+1}}\\
    &=x^{\sigma-d}\int\frac{\mathrm d\Omega}{(2\pi)^d}\int\mathrm dy\frac{e^{iy\cos\theta}}{y^{\sigma-d+1}}\\
    &\sim x^{\sigma-d}.
\end{aligned}
\end{equation}
The correlation function:
\begin{equation}
\begin{aligned}
    g(x)&=\langle\boldsymbol{\Psi}(\boldsymbol{x})\cdot\boldsymbol{\Psi}(0)\rangle\\
    &=\overline{\Psi}^2+\langle\Psi_T(\boldsymbol{x})\cdot\Psi_T(0)\rangle\\
    &=\overline{\Psi}^2+c\cdot x^{\sigma-d}.
\end{aligned}
\label{correlation function}
\end{equation}
Hence, for $\sigma<2$ at low temperatures, the correlation function takes on an algebraic form: $g(x)=g_0+c\cdot x^{-\eta_\ell}$, with $\eta_\ell=2-\sigma$. Note that the above derivation is applicable only at low temperatures; thus, we do not need to consider the renormalization of $K_2q^2$ terms as the system is far from criticality. Near or at the critical temperature, the system exhibits little or no spontaneous magnetization, and fluctuations play a significant role that cannot be considered negligible anymore.
For $\sigma=2$, however, the derivation above is not applicable, and the exact form of the correlation function remains unknown. In Eq.~\eqref{Hamiltonian}, two terms—$\frac{K_2}{2}q^2$ and $K_\sigma q^\sigma$—become degenerate, potentially leading to logarithmic behaviors. Next, we present the numerical result to verify the form of the correlation function—$g(x)=g_0+ c \cdot x^{-\eta_\ell}$—for $\sigma < 2$, and the logarithmic behavior of the system at $\sigma = 2$.

\textit{Numerical Observation}.
We have established that, at low temperatures for $\sigma < 2$, the correlation function takes the form:
$g(x) = g_0 + c \cdot x^{-\eta_\ell}$. While we aim to observe this behavior in our numerical results, directly analyzing the correlation functions poses challenges due to large finite-size corrections and the unknown constant $g_0$. Instead, the Fourier mode of magnetic susceptibility $\chi_k$ provides a more reliable means of analysis, as the constant term $g_0$ is effectively canceled out during the Fourier transformation. If the correlation function in Eq.~\eqref{correlation function} is valid, then $\chi_k$ should scale according to: $\chi_k \sim L^{2 - \eta_\ell}$. This scaling relation allows us to verify the system's behavior at low temperatures for $\sigma < 2$ and confirms whether the predicted power-law decay of correlations holds in our numerical simulations.

Up to the leading finite-size correction term, we propose the fitting ansatz of $\chi_k$ as,
\begin{equation}
    \chi_k=L^{2-\eta_\ell}(a_0+b_1L^{-\omega}).
\label{chik_fit}
\end{equation}
With $\eta_\ell$ fixed at $2-\sigma$, the $\chi_k$ data can be fitted well, and the results are presented in Fig.~\ref{chi_k}. The figure displays $\chi_k$ as a function of system size $L$ in a double-logarithmic plot for $\sigma = 1.25$, $1.75$, and $1.875$. In all cases, the data points for $\chi_k$ are close to a linear behavior, indicating the anticipated power-law relationship between $\chi_k$ and $L$. The slight upward curvature at smaller system sizes is attributed to finite-size correction terms. This linear trend in the large-$L$ regime demonstrates that the susceptibility follows the scaling relation $\chi_k \sim L^{2 - \eta_\ell}$ and thus confirms the correlation function in Eq.~\eqref{correlation function}. Additionally, the curves for different temperatures are almost parallel across all values of $\sigma$, and they can nearly be collapsed onto a single curve through vertical shifting. This observation indicates that the leading scaling exponent $\eta_\ell$ is temperature-independent.

At low temperatures, where spontaneous magnetization exists, according to the definition of $\xi$ in Eq.~\eqref{eq:2nd_moment_correlation}, $(\xi/L)^2$ is expected to follow the scaling relation: $(\xi/L)^2 \sim \frac{\langle M^2 \rangle}{\langle M_k^2 \rangle}$. Since both $\langle M^2 \rangle$ and $\langle M_k^2 \rangle$ are subject to similar finite-size corrections, their ratio is likely to cancel out the amplitudes of some communal correction terms, which may lead to a smaller finite-size effect. Considering the scaling form of $\langle M^2\rangle$ and $\langle M_k^2\rangle$, i.e. Eq.~\eqref{M2_scaling} and Eq.~\eqref{Mk2_scaling}, $(\xi/L)^2$ are fitted to the equation:
\begin{equation}
    (\xi/L)^2=a_0L^{\eta_\ell}+c.
\label{corrL_fit}
\end{equation}
In this approach, without considering any unknown correction terms, the data can be fitted well enough, indicating that $(\xi/L)^2$ indeed exhibits a smaller finite-size correction. The detailed fitting results and final estimates for $\eta_\ell$ are presented in Table~\ref{etal_detail} and Table~\ref{etal_results}, respectively. The results in Table~\ref{etal_results} are consistent with our conjecture that $\eta_\ell = 2 - \sigma$. While slight deviations are observed for $\sigma = 1.75$ and $\sigma = 1.875$ at $\beta = 1$, these are reasonable given the proximity to the critical point $\beta_c = 0.68380(7)$ and $0.70737(7)$. Our theoretical derivation is most applicable in the low-T regime, where the system is far from the critical point, and fluctuations are minimal compared to spontaneous magnetization. At even lower temperatures, specifically for $\beta = 2$, $4$, and $8$, all estimated values of $\eta_\ell$ are compatible with our derivation.

In the special case of $\sigma = 2$, where $\eta_\ell = 0$, we speculate that $\chi_k$ may exhibit a logarithmic scaling behavior, as $\chi_k \sim L^2\ln(L/L_0)^{\hat{\eta}_\ell}$, where $\hat{\eta}_\ell$ is the critical exponent of the logarithmic correction. To test this hypothesis, we attempt to fit $\chi_k$ using the following expression:
\begin{equation}
    \chi_k = a_0 L^2 /(\ln{L} + c_1)^{\hat{\eta}_{\ell}},
\label{LE2}
\end{equation}
where $a_0$, $L_0$, and $\hat{\eta}_{\ell}$ are unknown parameters. The results of these fits are shown in Table~\ref{LT2}. It is evident that, across different temperatures, the value of $\hat{\eta}_\ell$ from the fitting consistently approximates $1$. So we fix $\hat{\eta}_{\ell} = 1$ and re-fit $\chi_k$ to Eq.~\eqref{LE2}, with results also presented in Table~\ref{LT2}, supporting a possible scaling behavior: $\chi_k\sim L^2/\ln(L/L_0)$. This further implies that, at $\sigma = 2$, the real-space correlation function asymptotically saturates to a non-zero constant as an inverse logarithm of distance, $g(x)\sim g_0 + a\left(\frac{2\ln(x/b)-1}{\ln(x/b)^2} \right)$ for $x\gg 1$, where $a$ and $b$ are non-universal constants. This unusual inverse logarithmic decay of correlations highlights the nontrivial and marginal nature of the $\sigma = 2$ case. 

\begin{figure}
    \centering
    \includegraphics[width=\linewidth]{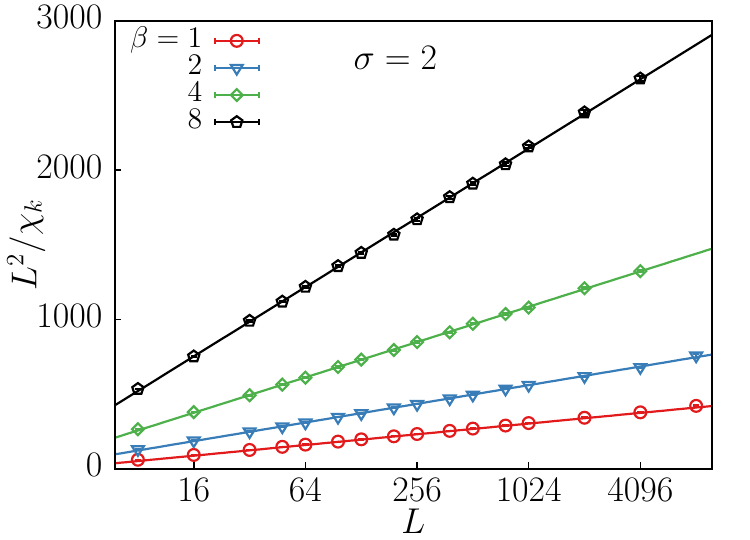}
    \caption{Demonstration of the logarithmic behavior for $\sigma=2$. $L^2/\chi_k$ is plotted in the semi-logarithmic coordinate as a function of the system size $L$ for different temperatures at $\sigma = 2$. The linearity of data indicates the logarithmic behavior: $\chi_k\sim L^2/\ln(L/L_0)$ in the low-T phase for the $\sigma = 2$ case.
    }
    \label{chik_sigma2}
\end{figure}

In Fig.~\ref{chik_sigma2}, we provide additional evidence by plotting $L^2/\chi_k$ as a function of $L$, with the $x$-axis in logarithmic coordinates. All the data points for different low temperatures show linear behavior, indicating $L^2/\chi_k \sim a \ln L + b$, which aligns with our assumption that $\chi_k \sim L^2/\ln(L/L_0)$.  {Moreover, from the fitting table~\ref{LT2}, one can find that the amplitude $a_0$ is proportional to the temperature $T$. Therefore, we propose that $\chi_k$ should satisfy $\chi_k= A(T)L^2/\ln(L/L_0)$ with $A(T)\propto T$ being an analytical function of $T$.} 

\begin{table*}[!ht]
    \centering
    \caption{Fits of $(\xi/L)^2$ to Eq.~\eqref{corrL_fit} for $\sigma=1.25$, 1.75 and 2. }
    \begin{tabular}{p{1.2cm}p{1.2cm}p{2cm}p{2cm}p{2cm}p{2cm}p{1.2cm}}
    \hline\hline
        $\sigma$&$\beta$ & $L_{\rm min}$ & $\eta_\ell$ & $a_0$ & $c$ & $\chi^2/$DF \\ \hline 
        1.25&1&32 & 0.751(1) & 0.357(4) & -0.46(3) & 10.9/11 \\ 
        &&48 & 0.750(2) & 0.361(5) & -0.50(5) & 9.8/10 \\ 
        &2&32 & 0.755(2) & 0.93(1) & -0.9(1) & 10.0/10 \\ 
        &&48 & 0.753(3) & 0.95(2) & -1.0(2) & 9.1/9 \\ 
        &4&48 & 0.750(4) & 2.13(6) & -2.6(5) & 7.9/9 \\ 
        &&64 & 0.751(5) & 2.10(7) & -2.3(8) & 7.6/8 \\ 
        &8&32 & 0.746(4) & 4.5(1) & -5.2(9) & 8.3/9 \\ 
        &&48 & 0.742(6) & 4.6(1) & -6(1) & 7.7/8 \\ \hline
        1.75&1&64 & 0.235(3) & 1.98(6) & -2.1(1) & 4.6/8 \\ 
        &&96 & 0.235(4) & 1.97(9) & -2.1(1) & 4.6/7 \\ 
        &2&32 & 0.246(2) & 5.2(1) & -5.6(2) & 5.5/10 \\ 
        &&48 & 0.247(3) & 5.1(1) & -5.5(2) & 5.4/9 \\ 
        &4&32 & 0.254(2) & 10.8(2) & -11.6(4) & 3.5/10 \\ 
        &&48 & 0.253(3) & 10.9(3) & -11.7(5) & 3.5/9 \\ 
        &8&32 & 0.250(4) & 24(1) & -26(1) & 8.1/10 \\ 
        &&48 & 0.254(5) & 23(1) & -24(2) & 7.0/9 \\ \hline
        1.875&1&32 & 0.093(3) & 6.1(3) & -6.4(4) & 7.2/9 \\ 
        &&48 & 0.100(4) & 5.5(3) & -5.7(3) & 3.9/8 \\ 
        &2&32 & 0.121(4) & 12.0(7) & -12.4(8) & 6.5/9 \\
        &&48 & 0.117(6) & 13(1) & -13(1) & 5.9/8 \\ 
        &4&32 & 0.123(7) & 27(2) & -28(3) & 10.4/9 \\ 
        &&48 & 0.12(1) & 29(4) & -30(4) & 9.4/8 \\ 
        &8&32 & 0.118(5) & 60(4) & -63(5) & 4.6/9 \\
        &&48 & 0.126(6) & 53(4) & -56(5) & 3.2/8 \\
        \hline\hline
    \end{tabular}
    \label{etal_detail}
\end{table*}

\begin{table*}[!ht]
    \centering
    \caption{Fits of $\chi_k$ to Eq.~\eqref{LE2} for $\sigma=2$ }
    \begin{tabular}{p{1.2cm}p{2cm}p{2cm}p{2cm}p{2cm}p{1.2cm}}
    \hline\hline
        $\beta$ & $L_{\rm min}$ & $\hat\eta_\ell$ & $c_1$ & $a_0$ & $\chi^2/$DF \\ \hline 
        1.0 &32 & 1.04(1) & -0.86(6) & 0.0213(8) & 6.6/9 \\
        &48 & 1.06(2) & -0.8(1) & 0.023(1) & 5.4/8 \\ 
        & 96 & 1 & -1.06(1) & 0.01918(7) & 4.0/7 \\ 
        &128 & 1 & -1.06(2) & 0.01919(9) & 4.0/6 \\ 
        2.0&48 & 0.99(2) & -0.8(1) & 0.0108(7) & 3.5/8 \\ 
        &64 & 0.98(3) & -0.8(1) & 0.011(1) & 3.5/7 \\
        &64 &1 & -0.72(1) & 0.01107(4) & 3.6/8 \\ 
        &96 &1 & -0.71(2) & 0.01109(5) & 3.5/7 \\
        4.0 & 32 & 1.03(2) & -0.5(1) & 0.0063(4) & 4.9/9 \\
        &64 & 1.04(6) & -0.4(3) & 0.007(1) & 4.6/7 \\ 
        &96 & 1.13(9) & 0.1(5) & 0.008(2) & 3.5/6 \\  
        &48 & 1 & -0.57(2) & 0.00587(2) & 5.4/9 \\ 
        &64 & 1 & -0.58(2) & 0.00585(3) & 4.9/8 \\ 
        8.0&48 & 1.10(7) & -0.1(3) & 0.0039(7) & 6.5/7 \\
        & 64 & 1.2(1) & 0.3(5) & 0.005(1) & 5.5/6 \\
        &32 &1 & -0.50(2) & 0.00300(1) & 10.1/9 \\ 
        &48 & 1&-0.53(3) & 0.00299(2) & 8.4/8 \\
        \hline\hline
    \end{tabular}
    \label{LT2}
\end{table*}

\begin{table*}[!ht]
    \centering
    \begin{threeparttable}
    \caption{Estimates of $\eta_\ell$ for different parameters are provided. The data in the `low-T' column combines the data from the previous four temperatures, where data in the `$\beta=1$' column for $\sigma=1.75, 1.875$ are discarded for their obvious deviation from other data, which is caused by their proximity to the critical temperature.}
    \begin{tabular}{p{2cm}p{2cm}p{2cm}p{2cm}p{2cm}p{2cm}p{2cm}}
    \hline\hline
        $\sigma$&$\beta=1$ & $\beta=2$ & $\beta=4$ & $\beta=8$ & low-T & theory \\ \hline
        1.25&0.751(3)&0.754(4)& 0.751(5) & 0.743(7) & 0.751(2)& 0.75\\
        1.75 & 0.237(6) & 0.247(3) & 0.253(3) & 0.250(4) & 0.250(2) & 0.25\\
        1.875 & 0.098(8) & 0.121(8) & 0.123(7) & 0.121(9) & 0.122(5) & 0.125\\
        \hline\hline
    \end{tabular}\label{LT5}
    \label{etal_results}
    \end{threeparttable}
\end{table*}

\subsection{$\sigma > 2$ cases}
\label{sigma_g2}

It is well-established that for $\sigma > 2$, the LR system belongs to the SR universality and exhibits QLRO at low temperatures. In this subsection, as a comparison to the $\sigma \le2$ case, we briefly demonstrate the low-T properties of the QLRO phase for $\sigma > 2$, using $\sigma = 3$ and $\sigma = 2.1$ as examples.

\begin{figure}
    \centering
    \includegraphics[width=\linewidth]{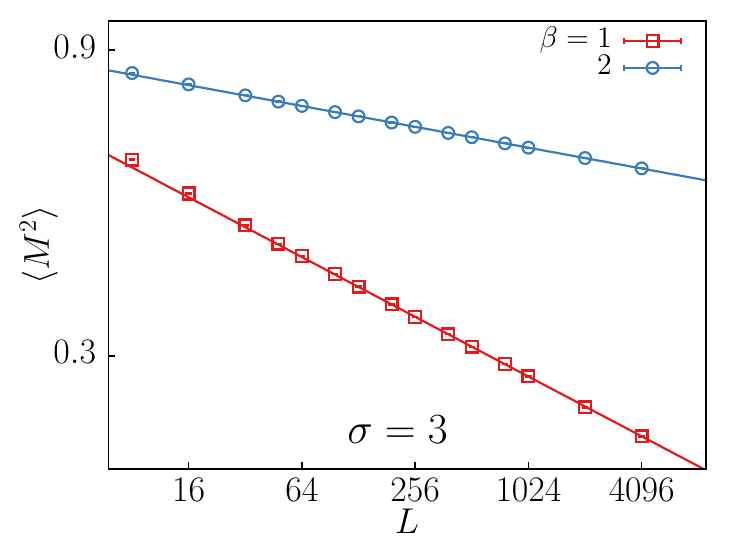}
    \caption{Demonstration of QLRO in the low-T phase for $\sigma=3$. The squared magnetization, $\langle M^2\rangle$, is plotted against $L$ using double-logarithmic coordinates. For both $\beta = 1$ and $\beta = 2$, the linearity of the data points indicates a power-law relationship between $\langle M^2\rangle$ and $L$, which is the signature of QLRO.}
    \label{M2_sigma3}
\end{figure}

Figure~\ref{M2_sigma3} plots $\langle M^2\rangle$ as a function of $L$ for $\sigma=3$ at double-log coordinates. The linearity of the data points indicates the scaling behavior $\langle M^2\rangle\sim L^{-\eta}$ and hence the correlation function's form: $g(r)\sim r^{-\eta}$. As $L\rightarrow \infty$, $\langle M^2\rangle$ reduces to 0, thus no spontaneous magnetization at low temperature. Moreover, the slope of the fitted line for the data of $\beta = 1$ is greater than that of $\beta = 2$. This suggests that the value of $\eta$ decreases as the temperature approaches 0. All these properties are typical signatures of QLRO. The case of $\sigma=2.1$ is not presented in Fig.~\ref{M2_sigma3} because it exhibits strong finite-size corrections due to the adjacency to $\sigma=2$. 

\begin{figure}
    \centering
    \includegraphics[width=\linewidth]{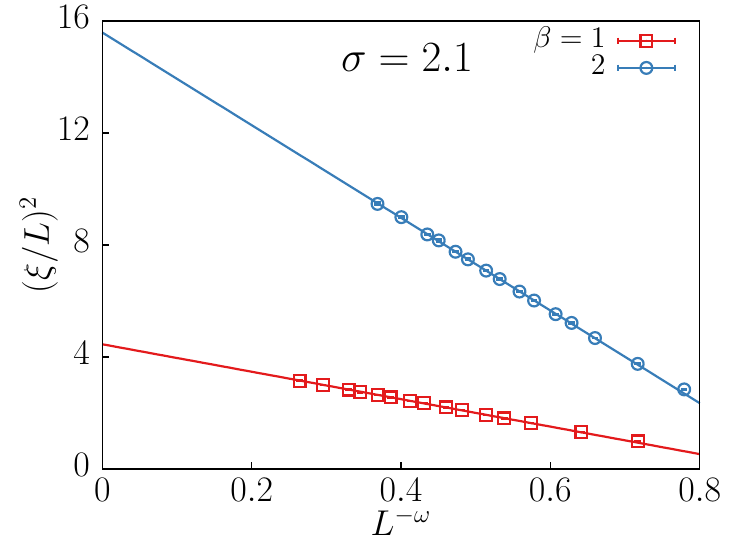}
    \caption{Fitting results of Eq.~\eqref{corrFit_sigma2.1} for $\sigma = 2.1$ at $\beta = 1$ and $\beta = 2$. The $x$-axis is $L^{-\omega}$, with $\omega = 0.16$ for $\beta = 1$ and $0.12$ for $\beta = 2$. The good agreement between the fitted line and the scatter points demonstrates the quality of the fit, and the $y$-intercept of the line provides the finite value of $(\xi/L)^2$ in the thermodynamic limit, suggesting the presence of QLRO.}
    \label{corr2_L}
\end{figure}

The QLRO nature of $\sigma=2.1$ can be revealed by analyzing the behavior of $(\xi/L)^2$. In Fig.~\ref{correL_lowT_s}, we have compared different behaviors of $(\xi/L)^2$ for varying $\sigma$. The diverging behavior for $\sigma\leq2$ indicates the existence of LRO, whereas the converging behavior shows the nature of QLRO. For $\sigma=2.1$, due to significant finite-size corrections, $(\xi/L)^2$ converges slowly, causing some ambiguity. To clearly demonstrate the converging behavior of $(\xi/L)^2$, we perform a fit on its finite-size data. According to Eq.~\eqref{eq:2nd_moment_correlation}, at low temperatures with $L$ large enough, $(\xi/L)^2$ has the scaling: $(\xi/L)^2\sim \langle M^2\rangle/\langle M_k^2\rangle$. For $\sigma>2$, $\langle M^2\rangle$ and $\langle M_k^2\rangle$ share the same leading scaling exponent $\langle M^2\rangle\sim\langle M_k^2\rangle\sim L^{-\eta}$, but may have different correction terms. Therefore, $(\xi/L)^2$ should converge to a constant following a series of power-law decays. Thus 
$(\xi/L)^2$ can be fitted to:
\begin{equation}
    (\xi/L)^2=a_0+a_1L^{-\omega}
\label{corrFit_sigma2.1}
\end{equation}
where $\omega$ is an unknown correction exponent and needs to be determined by fitting. The fitting results are shown in Fig.~\ref{corr2_L} with the largest system size up to $L=4096$. The linearity of the data points suggests that $(\xi/L)^2$ indeed scales in the form of Eq.~\eqref{corrFit_sigma2.1}. The finite extrapolated value indicates that $(\xi/L)^2$ converges to a finite constant in the thermodynamic limit, and hence the system enters a QLRO phase at low temperatures for $\sigma=2.1$.

\subsection{Revisiting the low-T properties of the crossover point $\sigma=2$}
\label{sigma_e2}

\begin{figure}[h]
    \centering
    \includegraphics[width=\linewidth]{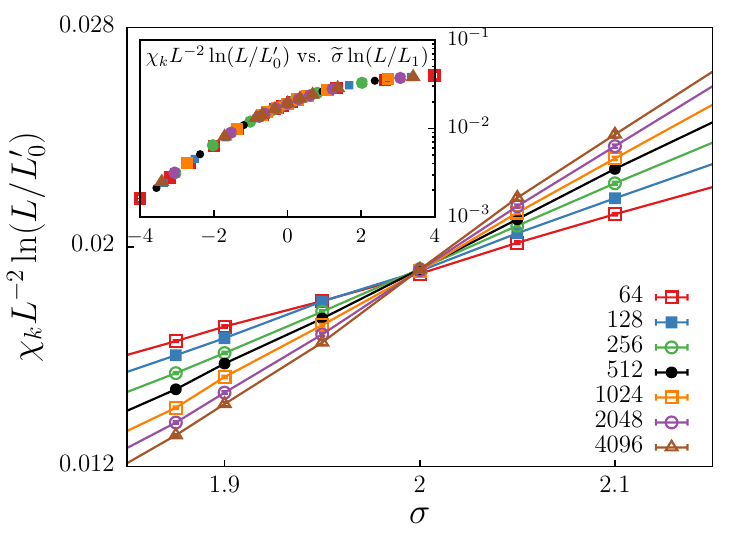}
    \caption{The semi-logarithmic plot of $\chi_kL^{-2}\ln(L/L'_0)$ as a function of $\sigma$ for several system sizes $L=64$ to 4096, with $y$-axis in logarithmic scale. The inset shows the data collapse of $\chi_kL^{-2}\ln(L/L'_0)$ vs. $\tilde{\sigma}\ln(L/L_1)$, where $\tilde{\sigma}= \sigma - 2$, $L'_0 = 2.89$ and $L_1 = 4.48 $.}
    \label{chik_s}
\end{figure}

\begin{figure}[h]
    \centering
    \includegraphics[width=\linewidth]{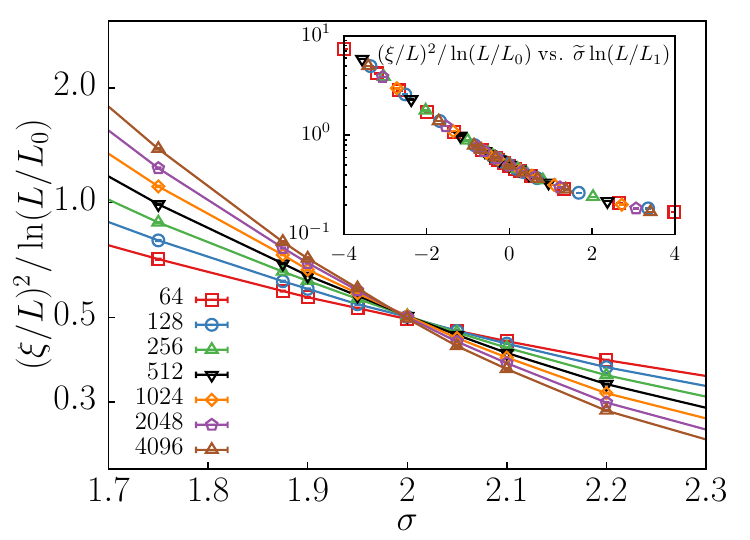}
    \caption{The semi-logarithmic plot of $(\xi/L)^2/\ln(L/L_0)$ as a function of $\sigma$ for several system sizes $L=64$ to 4096, with $y$-axis in logarithmic scale. The inset shows the data collapse of $(\xi/L)^2/\ln(L/L_0))$ vs. $\tilde{\sigma}\ln(L/L_1)$, where $\tilde{\sigma}= \sigma - 2$, $L_0 = 0.75$ and $L_1 = 4.48 $.}
    \label{corrL_s}
\end{figure}

The preceding subsections have established the existence of LRO for $\sigma\leq2$ and QLRO for $\sigma>2$ in the low-T phase. To gain deeper insights into the low-T properties of the system at the marginal point of $\sigma=2$,  we investigate the $2^{\text{nd}}$-moment correlation length $\xi$ as a function of $\sigma$ near $\sigma=2$. We fix the temperature at $T = 1.0$, a value that is below the critical temperature $T_c(\sigma = 2) = 1.3671(4)$. We begin by analyzing the scaling form of $\xi$ at $\sigma=2$. Section~\ref{Goldstone mode} has clearly demonstrated the logarithmic scaling behavior of $\chi_k$: $\chi_k\sim L^2/\ln(L/L_0)$ at $\sigma=2$ in the low-T phase. Given the definition Eq.~\eqref{eq:2nd_moment_correlation} and considering the finite magnetization in the low-T phase, the $2^{\text{nd}}$-moment correlation length $\xi$ should obey a similar logarithmic scaling form: $\xi\sim L\sqrt{\frac{\langle M^2\rangle}{\langle M_k^2\rangle}}\sim L\ln(L/L_0)^{\frac{1}{2}}$. 
To explore the transition across $\sigma=2$, we analyze how the behavior of $\chi_k$ and $\xi$ as a function of $\sigma$.

Figure~\ref{chik_s} and \ref{corrL_s} presents a detailed finite-size scaling analysis of $\chi_kL^{-2}\ln(L/L'_0)$ and $(\xi/L)^2/\ln(L/L_0)$ as a function of $\sigma$ for several system sizes $L$. A key feature is the clear crossing of curves at $\sigma = 2$, signifying a change between LRO and QLRO. To capture the behavior near $\sigma = 2$, we propose a refined scaling form for $\chi_k$ and $\xi$, incorporating logarithmic corrections at $\sigma = 2$ in the low-T phase $T < T_c$. Specifically, we conjecture that $\chi_k$ scales as:
\begin{equation}
    \chi_k = L^{2} \ln(L/L'_0)^{-1} \chi_k'(\tilde{\sigma} \ln\left(L/L_1\right)) ,
    \label{chi_FSS}
\end{equation}
and $\xi$ scales as:
\begin{equation}
    \xi = L \ln(L/L_0)^{\frac{1}{2}} \xi'(\tilde{\sigma} \ln\left(L/L_1\right)) ,
    \label{xi_FSS}
\end{equation}
where $\tilde{\sigma} = \sigma - 2$ represents the deviation from $\sigma_*$; $\chi'_k$ and $\xi'$ are universal scaling functions that govern the behavior near $\sigma = 2$; $L_0$, $L'_0$ and $L_1$ are constants determined via fitting. The insets of Fig.~\ref{chik_s} and \ref{corrL_s} demonstrate the efficacy of the scaling hypotheses by respectively showing a collapse of the scaled $\chi_k$ and $\xi$ data points onto single, universal curves. The data collapse provides strong numerical support for conjectured scaling forms and further confirms that $\sigma = 2$ represents a marginal point where the system undergoes a transition between the LRO and QLRO phases at low temperatures.

\section{Power-law divergence of correlation length at high temperature}\label{sec:results_II}

\begin{figure}[ht]
    \centering
    \includegraphics[width=\linewidth]{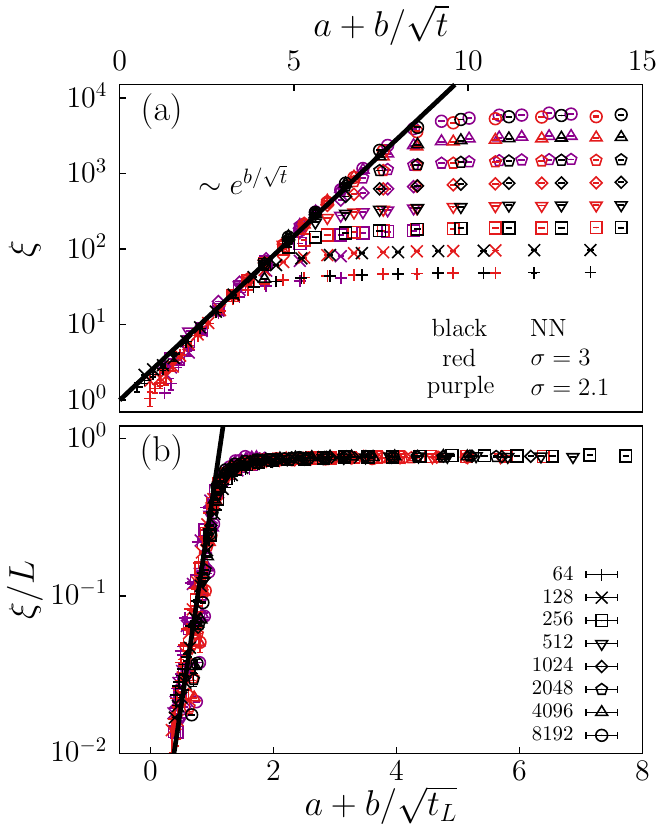}
    \caption{Demonstration of the exponential growth of $\xi$ and corresponding FSS, indicating the BKT transition for $\sigma > 2$.
    (a) The semi-logarithmic plot of $\xi$ as a function of $a + b / \sqrt{t}$ for $\sigma = 2.1$ (purple), 3 (red) and NN case (black), where $t = (T-T_c)/T_c$ and the non-universal constants are $a = 0.69, -0.22, -1.20$ and $b = 0.6, 1.1, 1.7$, respectively. 
    (b) The semi-logarithmic plot of the ratio $\xi/L$ as a function of $a + b/\sqrt{t_L}$, where $ t_L = t [\ln (L / L_0) ]^2$ with the non-universal length scale simply set by $L_0 = 1$. Here, the the non-universal constants are $a = 0, 0.12, 0.25$ and $b = 0.6, 1.1, 1.7$,  respectively.}
    \label{xi_chi_t_s_l_2}
\end{figure}

\begin{figure}[ht]
    \centering
    \includegraphics[width=\linewidth]{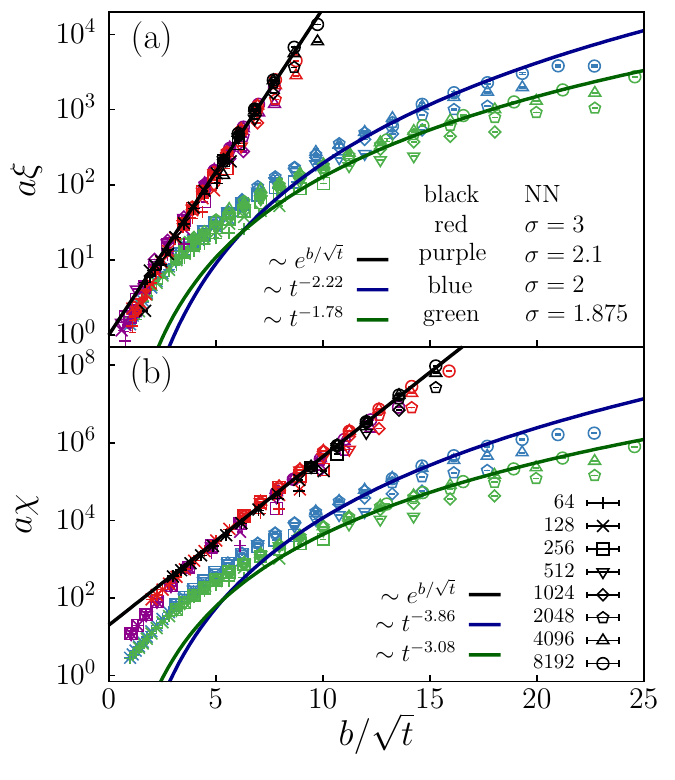}
    \caption{Deviation of the $2^{\text{nd}}$-moment correlation length $\xi$ and susceptibility $\chi$ growth at $\sigma = 2$ from the BKT scaling. Panel (a) shows a semi-logarithmic plot of $a\xi$ as a function of $b/\sqrt{t}$ for various $L$ at $\sigma=1.875$ (green), 2 (blue), 2.1 (purple), 3 (red), and the NN case (black), where the reduced temperature is $t = (T - T_c)/T_c$, and $a = 1, 1, 0.5, 1.25, 3.33$ and $b = 1, 1,  0.6, 1.1, 1.7,$ respectively. Panel (b) shows a semi-logarithmic plot of $a\chi$ as a function of $b/\sqrt{t}$ for various $L$ at $\sigma=1.875$ (green), 2 (blue), 2.1 (purple), 3 (red), and the NN case (black), where $a = 1, 1, 4, 23.53, 66.67$ and $b = 1, 1, 1.05, 2, 3,$ respectively.
    For the $\sigma = 2.1, 3,$ and NN case, the black solid line, which scales as $e^{b/\sqrt{t}}$, serves as a guide to the eye for the exponential growth of $\xi$ and $\chi$, characterizing the BKT transition. However, for $\sigma = 1.875$ and $2$, the dark-green and dark-blue curved lines, scaling as $t^{-1.78}$ and $t^{-2.22}$ in panel (a), and $t^{-3.08}$ and $t^{-3.86}$ in panel (b), serve as guides to the eye for the power-law growth of $\xi$ and $\chi$, indicating a second-order phase transition.}
    \label{xi_chi_t}
\end{figure}

\begin{figure}[ht]
    \centering
    \includegraphics[width=\linewidth]{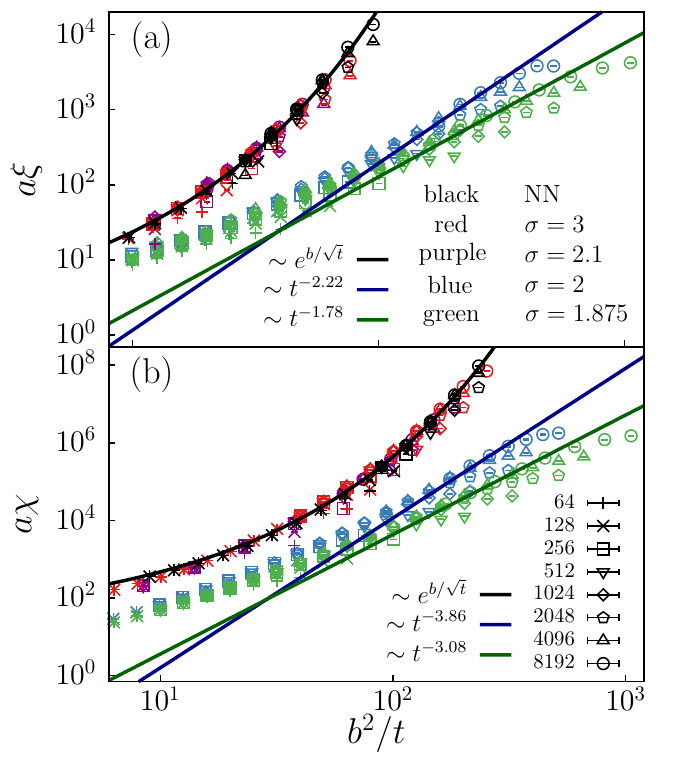}
    \caption{Growth of the $2^{\text{nd}}$-moment correlation length $\xi$ and susceptibility $\chi$ growth at various $\sigma$ in double-logarithmic scale. Panel (a) shows $a\xi$ as a function of $b^2/t$ for various $L$ at $\sigma=1.875$ (green), 2 (blue), 2.1 (purple), 3 (red), and the NN case (black), where the reduced temperature is $t = (T - T_c)/T_c$. Panel (b) shows $a\chi$ as a function of $b^2/t$ for various $L$ at $\sigma=1.875$ (green), 2 (blue), 2.1 (purple), 3 (red), and the NN case (black). The values of $a$ and $b$ are the same as those in Fig.~\ref{xi_chi_t}.
    For $\sigma = 1.875$ and $2$, $\xi$ and $\chi$ asymptotically follow a power-law growth, indicating a second-order phase transition. As a comparison, for $\sigma = 2.1, 3,$ and NN cases, the exponential growth of $\xi$ and $\chi$ for a BKT transition can be observed (the black curve).
    }
    \label{xi_chi_t_dou_log}
\end{figure}

In this section, we focus on the physical properties of the high-T phase as the system approaches the critical point at $\beta_c$ from the disordered phase. Specifically, we investigate the growth of the second-moment correlation length $\xi$ and susceptibility $\chi$. The growth form of $\xi$ and $\chi$ depends on the type of the transition, providing a clearer indication of universality than critical exponents and allowing one to identify the threshold between SR and LR universality. The critical points \( \beta_c \) for \( \sigma = 1.875, 2, 2.1, \) and \( 3 \) are determined through FSS analysis, whose value can be found in Table~\ref{Kc_exponent}; for the NN case, previous numerical investigation gives \( \beta_c = 1.11996(6) \)~\cite{komura2012a}.

\begin{figure*}
    \centering
    \includegraphics[width=0.85\linewidth]{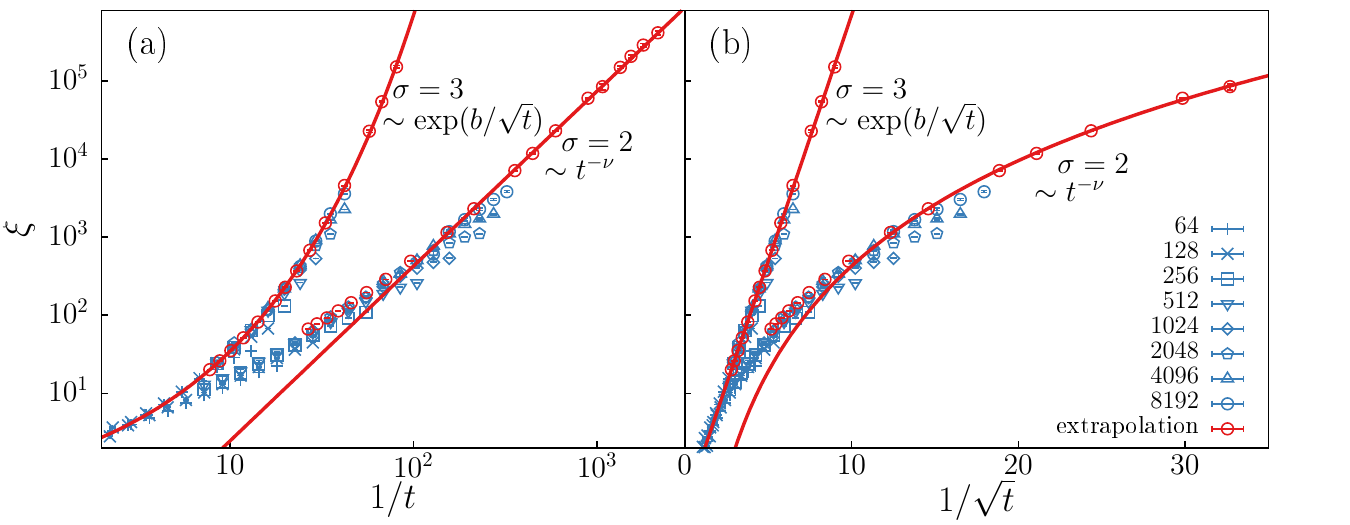}
    \caption{The extrapolation results of $\xi$ along with the original data for $\sigma=2,3$. The blue dots represent the original data obtained directly from the simulation without extrapolation, as already shown in Fig.~\ref{xi_chi_t} and \ref{xi_chi_t_dou_log}; the red dots represent the extrapolated value. Panel (a) is plotted in a double-log coordinate. The distinct linearity of the extrapolated data for $\sigma=2$ demonstrates the power-law divergence of $\xi$: $\xi\sim t^{-\nu}$ with $\nu\approx2.24$. Such power-law divergence of $\xi$ is the signature of a second-order phase transition. For comparison, in panel (b), $\xi$ is plotted versus $1/\sqrt{t}$ in a semi-log coordinate. The linearity of the extrapolated data for $\sigma=3$ demonstrates the exponential divergence of $\xi$: $\xi\sim\exp(b/\sqrt{t})$, which is the signature of a BKT phase transition.}   
    \label{extrapolation}
\end{figure*}

As the system approaches a BKT transition from a disordered phase, $\xi$ and $\chi$ diverge exponentially as \( \sim \exp(b/\sqrt{t}) \)~\cite{J_M_Kosterlitz_1973, J_M_Kosterlitz_1974}, where \( t = (T - T_c)/T_c \) is the reduced temperature, and \( b \) is a non-universal constant. As shown in Fig.~\ref{xi_chi_t_s_l_2}(a), a semi-logarithmic plot of \( \xi \) versus \( a + b\sqrt{t} \) is presented for \( \sigma = 2.1 \) (purple), \( 3 \) (red), and the NN case (black), where the non-universal constant \( a \) comes from the amplitude of the scaling \( \sim \exp(b/\sqrt{t}) \), and the specific values of \( a \) and \( b \) are provided in the corresponding caption.

For a given $t$ that is not so close to the critical point, the growth of $\xi$ as a function of $t$ then reveals the thermodynamic growth law of correlation length. However, when the system enters the finite-size critical window, where $\xi \sim L$, the $\xi$ curve begins to bend to a plateau due to finite-size scaling behavior, thus deviating from the thermodynamic growth law. 
(Note that this analysis is different from the FSS analysis. Here, one first takes the $L\rightarrow \infty$ limit for a given temperature $T$ and then takes the $T \rightarrow T_c$ limit to examine the asymptotic growth law of $\xi$. In contrast, the FSS analysis first takes the $T \rightarrow T_c$ limit for a given $L$, followed by the $L\rightarrow \infty$ limit, studying the scaling behavior within the finite-size critical window.)
As shown in Fig.~\ref{xi_chi_t_s_l_2}(a), the straight black line in the semi-log plot clearly demonstrates the exponential growth of $\xi$ for $\sigma > 2$, the typical BKT scaling behavior. In the large $t$ regime (the bottom left corner of Fig.~\ref{xi_chi_t_s_l_2}(a)), where the system is far from criticality, $\xi$ slightly deviates from the exponential growth.
Moreover, in Fig.~\ref{xi_chi_t_s_l_2}(b), we further plot the ratio $\xi/L$ versus $b/\sqrt{t_L}=b/\sqrt{t(\ln L / L_0)^2}$ to examine the FSS of the expoential growth, where $L_0$ is non-universal length scale, simply set as $1$. The scaling field $t_L=t[\ln (L/L_0)]^2$~\cite{PhysRevB.88.104427, PhysRevB.55.R11949} originates from the exponential divergence of the correlation length near the BKT transition $\xi \sim L \sim \exp(b/\sqrt{t})$. The data points collapse onto a single curve, showing that for $\sigma > 2$, the phase transitions belong to the BKT universality class. For $\sigma=2.1$, taking into account that it is so close to the crossover point $\sigma=2$, it is somewhat surprising that its $\xi$ data can collapse well onto the curve for $\sigma=3$ and NN.

On the other hand, near a second-order transition, $\xi$ diverges algebraically as $\xi \sim t^{-\nu}$, with $\nu$ as the correlation length exponent, and $\chi$ diverges algebraically as $\chi \sim t^{-\gamma}$, with $\gamma$ as the susceptibility exponent. Hence, by investigating the growth of $\xi$ and $\chi$ near $\beta_c$, we can readily reveal the type of transition at different $\sigma$, particularly the transition at $\sigma = 2$. 

\begin{figure*}[!ht]
    \centering
    \includegraphics[width=\linewidth]{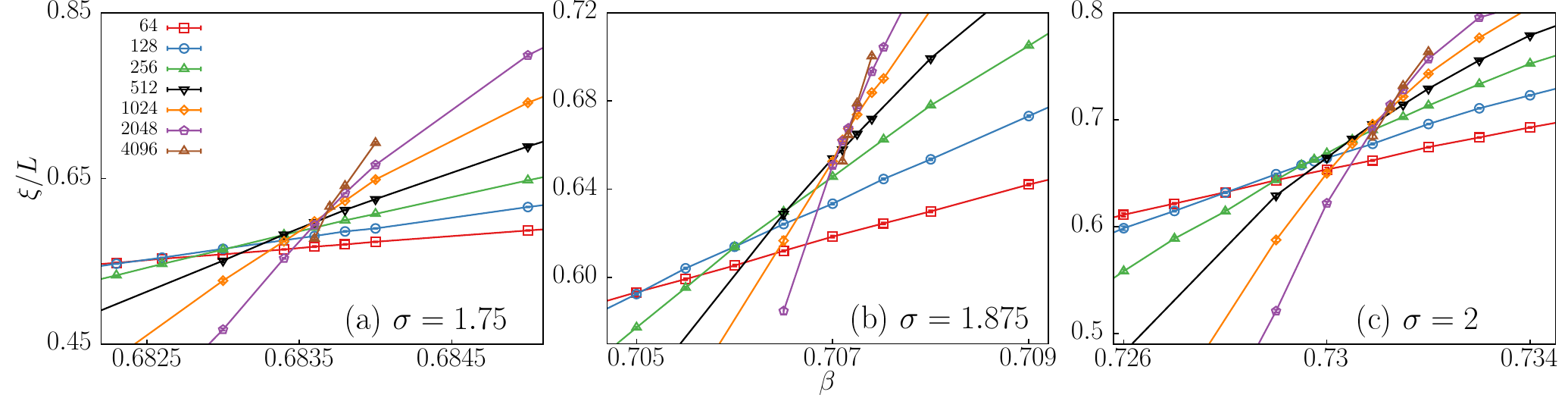}
    \caption{The zoom-in of the second-moment correlation length ratio, $\xi/L$, as a function of inverse temperature $\beta$ for various system sizes at $\sigma$: $1.75$ (a), $1.875$ (b), and $2$ (c). Clear crossing behavior can be observed in all the cases, despite the shift of the crossing point due to the finite-size effect.
    }
    \label{Q_crossing_zoom}
\end{figure*}

In Fig.~\ref{xi_chi_t}(a), a semi-logarithmic plot of $a\xi$ versus $b/\sqrt{t}$ for various $\sigma$ is presented. The non-universal constants $a$ and $b$, obtained through the fitting, enhance the overall readability of plots without affecting the universal growth law. The specific values of $a$ and $b$ are provided in the corresponding caption. For $\sigma = 2.1$ (purple), $\sigma = 3$ (red), and the NN case (black), the straight black line demonstrates that the correlation length diverges exponentially $\xi \sim \exp(b/\sqrt{t})$, characterizing a typical BKT transition. For clarity, we exclude certain data points in the finite-size critical window, which manifests as a plateau.
For $\sigma=1.875$ (green) and $2$ (blue), when sufficiently away from $\beta_c$, i.e. $b/\sqrt{t}$ is small, $\xi$ behaves seemingly like that for $\sigma = 2.1$, $3$ and the NN case. However, near the critical point, i.e., as $b/\sqrt{t}$ approaches infinity, the behavior of $\xi$ becomes increasingly different from SR cases ($\sigma > 2$) and deviates from the exponential growth, suggesting a different universality class. In fact, the growth of $\xi$ for $\sigma=1.875$ and $2$ can be approximately described by a power-law function (the curved lines in dark green and dark blue), indicating a second-order transition. 
In Fig.~\ref{xi_chi_t}(b), a similar analysis is applied to the susceptibility, $\chi$. The plot shows $a\chi$ versus $b/\sqrt{t}$ on a semi-logarithmic scale. For the BKT cases ($\sigma = 2.1$, $3$, and NN), $\chi$ diverges exponentially, with the straight black line guiding the eye. On the other hand, for $\sigma = 1.875$ and $2$, $\chi$ follows a power-law growth near the critical point, with the dark-blue and dark-green curved lines guiding the eye, suggesting a second-order phase transition.

Figure.~\ref{xi_chi_t_dou_log} presents the log-log plot of \( a\xi \) and \( a\chi \) versus \( {b^2}/{t} \), where the x-axis is the square of \( {b}/{\sqrt{t}} \). In Fig.~\ref{xi_chi_t_dou_log}(a), for \( \sigma = 1.875 \) and \( 2 \), the growth of $\xi$ asymptotically follows a linear relation as \( {b^2}/{t} \rightarrow \infty \) (the straight dark-green and dark-blue lines), consistent with the power-law growth behavior of a second-order transition. The deviation of $\xi$ from linearity and the discrepancy between the exponent estimated from Fig.~\ref{xi_chi_t_dou_log}(a) and the fitting result in Table~\ref{Kc_exponent_} are likely due to the strong correction to scaling from the irrelevant fields. Specifically, $\xi\sim t^{-\nu} g(u/t^{y_u/y_t})$, where $u$ is some irrelevant field and $y_u < 0$ is the correpsonding scaling exponent. Nevertheless, the growth behavior of $\xi$ is already capable of clearly revealing the type of phase transition for various $\sigma$. For the growth of \( \chi \) in Fig.~\ref{xi_chi_t_dou_log}(b), the straight dark-green and dark-blue lines illustrate a power-law behavior, as \( \chi \sim t^{-\gamma} \) for \( \sigma = 1.875 \) and \( 2 \), again indicating a second-order transition. 

The growth behavior of $\xi$ for $\sigma=2, 1.875$ slightly deviates from a power law in Fig.~\ref{xi_chi_t_dou_log}(a), primarily because of the strong correction from the irrelevant fields. To mitigate such corrections, one needs to observe a larger $\xi$ closer to the critical point. However, brute-force simulations at larger systems are numerically challenging and impractical. Instead, one can exploit the FSS behavior of $\xi$ and extrapolate $\xi$ to obtain its thermodynamic value. The extrapolation method has been successfully employed to the classical 2D Heisenberg model~\cite{PhysRevLett.70.1735, PhysRevLett.75.1891, PhysRevLett.74.2969, Yao2025}. This method is based on the FSS ansatz:
\begin{equation}
    \frac{\mathcal{\xi}(\beta,sL)}{\mathcal\xi(\beta,~L)}=F_\mathcal{\xi}\left(\frac{\xi(\beta,L)}{L};s\right)+O(\xi^{-\omega},L^{-\omega})
    \label{extrapolation_eq}
\end{equation}
where $s$ is a fixed rescaling factor, typically set at $2$ for the convenience of MC simulations; $F_\xi$ is a universal function of $\xi/L$, depending on the rescaling factor $s$; the second term $O(\xi^{-\omega}, L^{-\omega})$ contains further correction terms, which vanishes when $\xi$ and $L$ becomes large enough. In 
the ansatz~\eqref{extrapolation_eq}, $F_\xi(x)$ typically takes the form:
\begin{equation}
    F_\mathcal{\xi}(x)=1+a_1e^{-1/x}+a_2e^{-2/x}+\cdots+a_ne^{-n/x}.
\label{extra_fit}
\end{equation}
with $n$ usually no more than $12$~\cite{Yao2025}. The correction term $O(\xi^{-\omega}, L^{-\omega})$ is negligible in the 2D Heisenberg case, which is also the case for the 2D LRXY model at $\sigma=3$. However, for $\sigma=2$, because of large finite-size correction, $O(\xi^{-\omega},L^{-\omega})$ term in ansatz~\eqref{extrapolation_eq} can not be ignored. The fitting equation for ansatz~\eqref{extrapolation_eq} is thus adjusted to:
\begin{equation}
    \frac{\mathcal{\xi}(\beta,sL)}{\mathcal\xi(\beta,~L)}=1+a_1e^{-1/x}+a_2e^{-2/x}+\cdots+a_ne^{-n/x}+bL^{-\omega}.
    \label{extra_fit2}
\end{equation}
The fitting results for $\sigma=2$ and $3$ are presented in Table~\ref{extra_fit2_detail} and \ref{extra_fit_detail} in Appendix~\ref{appen_A}. With the concrete form of Eq.~\eqref{extrapolation_eq}, one can extrapolate the truncated $\xi$ to its thermodynamic value. See Ref.~\cite{Yao2025} for details of the method. The final extrapolation results, along with original data in Fig.~\ref{xi_chi_t} and \ref{xi_chi_t_dou_log}, are shown in Fig.~\ref{extrapolation}. The extrapolated $\xi$ reveals the characteristic signatures of the phase transition more clearly. In Fig.~\ref{extrapolation}(a), $\xi$ is plotted vs. $1/t$ in the double-log scale. The linear behavior of the extrapolated data for $\sigma=2$ demonstrates a clear power-law divergence of $\xi$ with $\nu\approx 2.24$, characteristic of a second-order phase transition. In Fig.~\ref{extrapolation}(b), $\xi$ is plotted vs. $1/\sqrt{t}$ in the semi-log scale, and the linearity of the extrapolated data for $\sigma=3$ demonstrates the exponential divergence of $\xi$, consistent with a BKT phase transition. Moreover, the extrapolated correlation lengths agree well with the original simulation data for both the $\sigma=2$ and $3$ cases, affirming the robustness and reliability of our extrapolation procedure.

In all, our analysis of the growth behavior of $\xi$ and $\chi$ strongly suggests that the system undergoes a second-order transition for $\sigma \le 2$, while a BKT transition for $\sigma > 2$; hence, the threshold between SR and LR universality is at $\sigma_* = 2$.

\section{Critical properties} \label{sec:results_III}

\begin{figure*}[th]
    \centering
    \includegraphics[width=\linewidth]{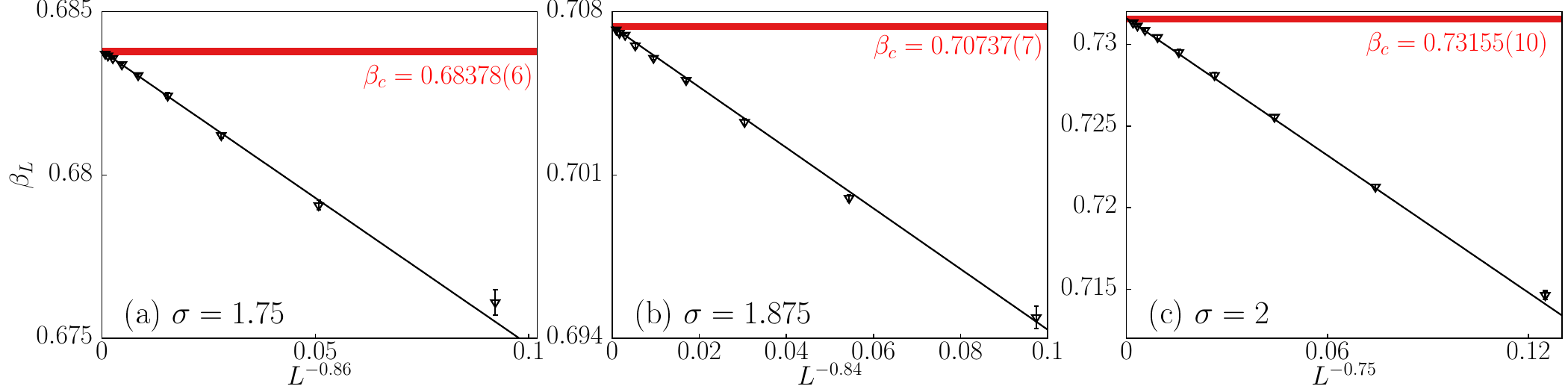}
    \caption{The extrapolation of $\xi/L$ curves' crossing point $\beta_L$ of different system sizes. The $x$-axis represents the correction term $L^{-\omega}$: $L^{-0.86}$ in panel (a), $L^{-0.84}$ in panel (b), and $L^{-0.75}$ in panel (c); the $y$-axis denotes the inverse temperature $\beta$ at which the correlation length ratios for two consecutive system sizes intersect. The red line indicates the extrapolated value of $\beta$ as the system size approaches infinity.}
    \label{crossing_point}
\end{figure*}

In the previous sections, we have studied the low-T and high-T properties of the system for $\sigma\le 2$. In this section, we focus on the system's critical properties for $\sigma\leq2$. The fitting of the critical point and critical exponent is elaborated in subsection~\ref{critical_point_fit} and \ref{critical_exponent_fit}, respectively. Then, to determine the reliability of the estimate of $\eta$, the correlation function is also studied in subsection~\ref{double_power}. Finally, we take a brief look at the properties of the specific-heat-like quantity in subsection~\ref{Cv}, which provides additional hints of the crossover by showing distinct behaviors in the LR and SR universality.

\subsection{The Fit of Critical Point}
\label{critical_point_fit}

In this subsection, we focus on determining the critical temperature for various values of $\sigma$. For $\sigma \leq 2$, a second-order phase transition from a disordered phase to a ferromagnetic phase happens, where the dimensionless ratio curves, such as $Q_m$ and $\xi/L$, for different system sizes, should intersect at the critical point. Figure~\ref{Q_crossing_zoom} presents $\xi/L$ as a function of inverse temperature $\beta$ for system sizes ranging from $L = 64$ to $L = 4096$, and for $\sigma = 1.75$, $1.875$, and $2.0$. Unlike Fig.~\ref{Q_crossing}, which covers a broader range, this figure zooms in on a small regime of $\beta$ to observe the crossing behavior of $\xi/L$ for different system sizes. However, Fig.~\ref{Q_crossing_zoom} shows that the crossing points for different system sizes do not intersect at a single value of $\beta$. Instead, these points shift toward higher values of $\beta$ as the system size increases, a phenomenon attributed to finite-size effects. The distance between the crossing points decreases as the system size grows, and with larger system sizes, the crossing points eventually converge at a single value of $\beta$, marking the critical point. To confirm this behavior and accurately determine the critical point, a finite-size analysis is performed at the crossing point $\beta_L$ between system sizes of $L$ and $L/2$.

\begin{figure*}
    \centering
    \includegraphics[width=\linewidth]{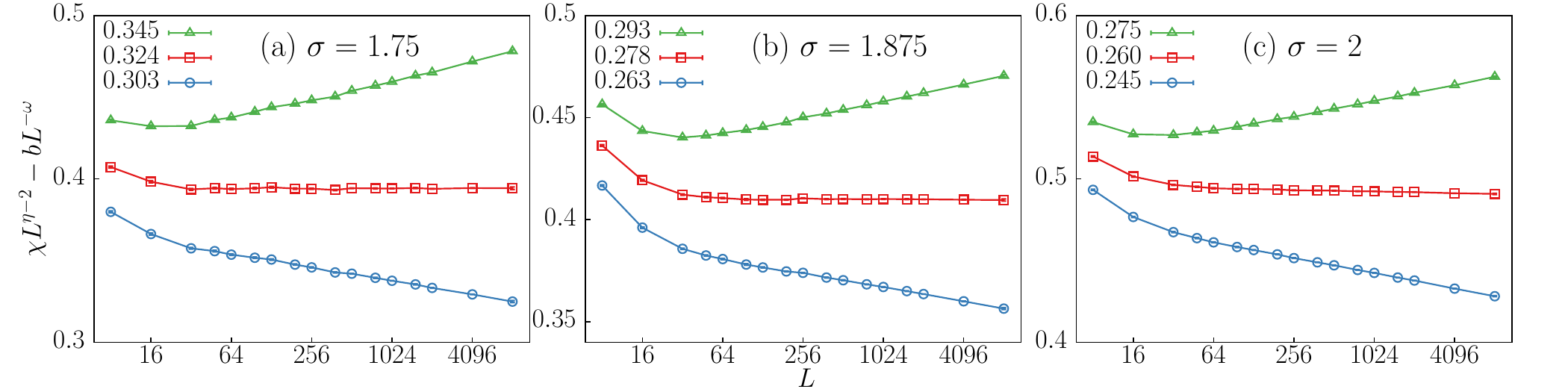}
    \caption{Demonstration of the reliability of the fitting of $\chi$. The semi-logarithmic plots of $\chi L^{\eta - 2} - bL^{-\omega}$ versus $L$ for $\sigma = 1.75, 1.875,$ and $2$ are shown (where $\omega=0.48,0.42,0.5$ respectively).  {The red dots represent $\eta$ at the fitted center value $\eta_c$, while the green and blue ones represent $\eta$ deviating from the center value by positive and negative three standard deviations, respectively. The red square approaches a horizontal line as $L \rightarrow \infty$, while others show evident growing and decreasing trend, indicating the precise estimate for $\eta$ that we obtain.}}
    \label{chi_appro_c}
\end{figure*}

\begin{figure*}
    \centering
    \includegraphics[width=\linewidth]{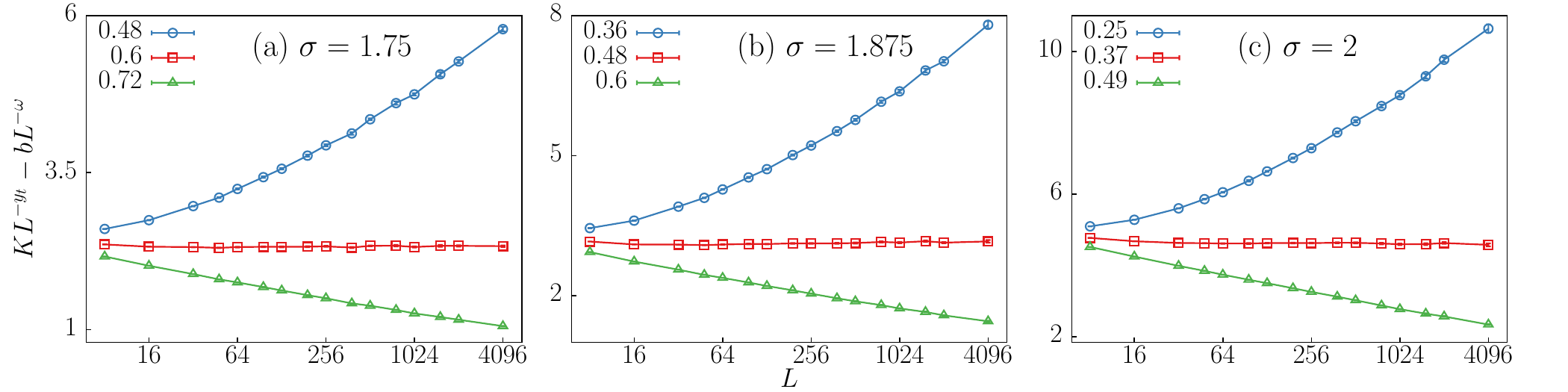}
    \caption{Demonstration of the reliability of the fitting of $K$. The semi-logarithmic plots of $K L^{-y_t} - bL^{-\omega}$ versus $L$ for $\sigma = 1.75, 1.875,$ and $2$ are shown (where $\omega=0.25,0.25,0.2$ respectively).  {The red dots represent $y_t$ at the fitted center value, while the blue and green ones represent $y_t$ deviating from the center value by positive and negative three standard deviations, respectively. The red square approaches a horizontal line as $L \rightarrow \infty$, while others show evident growing and decreasing trend, indicating the precise estimate for $y_t$ that we obtain (note that $y_t=1/\nu$).} It is worth noting that, compared to $\chi$, since $K$ is defined by the ratio of two vanishing covariance and variance in Eq.~\eqref{K_define}, it exhibits larger errors. This is particularly evident for $L = 8192$, where the difficulty in simulating longer Markov chain lengths results in significant errors for $K$ under this parameter. Therefore, the data point for $L = 8192$ is not included  in the fitting of $K$.}
    \label{K_appro_c}
\end{figure*}

The final crossing point, or critical inverse temperature $\beta_c$, is obtained by fitting $\beta_L$ to the following equation:
\begin{equation}
\beta_L = \beta_c + aL^{-\omega},
\end{equation}
where $\omega$ is related to the correlation length exponent $\nu$ and the leading irrelevant exponent in the renormalization group framework. Figure~\ref{crossing_point} illustrates the fitting results, with the $x$-axis representing $L^{-\omega}$ and the $y$-axis showing the corresponding values of $\beta_L$. In each case, the fitting lines exhibit finite intercepts on the $y$-axis, highlighted by red stripes. This confirms that the $\xi/L$ curves finally intersect at a single critical point as the system size increases. The intercepts provide an estimate of the critical temperature $\beta_c$. For further accuracy, the dimensionless ratios, such as $\xi/L$ and $Q_m$, are also fitted with respect to $L$ and $\epsilon=\beta - \beta_c$ near the critical point to refine the estimate of the critical point.
Considering the universal part of dimensionless ratio $Q$ as in Eq.~\eqref{eq:Qm_scale} and \eqref{eq:corrL_scale}, Certain Taylor expansion provides a fitting equation of $Q$:
\begin{equation}
\begin{aligned}
    Q(\epsilon,u,L)&=  a_0+\sum_{i=1}^{m}a_i(\epsilon L^{y_t})^i+b_1 L^{-\omega} +b_2 L^{-2\omega} \\
    &+ c_1\epsilon L^{y_t-\omega} + c_2\epsilon^2L^{2y_t-\omega},
\end{aligned}
\label{Q_fiteq}
\end{equation}
where the value of $m$ depends on the specific fitting process and is typically taken as $2$ or $3$; $L^{-\omega} $ term comes from the irrelevant field $u$, and $\omega=-y_u$. In the formula above, terms such as $\epsilon^2L^{y_t}, \epsilon^3L^{y_t}, \epsilon^3L^{2y_t}, \cdots$, are not under consideration since including such terms in the specific fitting process has almost no impact on the fit, and their amplitudes are also very close to zero. For the fit of $Q$, when $\sigma \leq 1.5$, $Q_m$ is more suitable for fitting, whereas when $1.5<\sigma \leq 2$, $\xi/L$ provides a better fit.
By combining the two quantities, relatively accurate estimates of critical points can be obtained, and they are consistent with those derived from the fit of $\beta_L$. The final estimated values of $\beta_c$ are presented in Table~\ref{Kc_exponent} (Note that, except for $\sigma=1.25,1.5,1.75,1.875$ and $2$, the critical points under other parameters are only determined with much less precision since they are mainly used for drawing the phase diagram in Fig. \ref{PD}).

\begin{table}[!ht]
    \centering
    \caption{Estimations of critical point $\beta_c$ for various $\sigma$ shown in Fig.~\ref{PD}.  {Because of the normalization procedure in Eq.~\eqref{normalization}, when $\sigma\to-2$, $\beta_c\to\frac12$, the critical point of XY model on the complete graph; when $\sigma\to\infty$, $\beta_c$ approaches the value of the NN case.}}
    \begin{threeparttable}
    \begin{tabular}{l|l||l|l}
    \hline\hline
        ~~~$\sigma$ & ~~~~~~$\beta_c$ & ~$\sigma$ & ~~~~~~$\beta_c$ \\ \hline
        0.25  & 0.503(2)      & 2     & 0.7315(2)       \\
        0.45  & 0.5075(2)     & 2.1   & 0.753(2)         \\
        0.65  & 0.524(2)      & 2.2   & 0.783(2)       \\
        0.85  & 0.545(2)      & 2.5   & 0.835(2)       \\
        1     & 0.5654(1)     & 3     & 0.899(5)         \\
        1.25  & 0.599615(6)   & 3.5   & 0.950(5)     \\
        1.5   & 0.63936(1)    & 4     & 0.980(5)   \\
        1.75  & 0.68380(7)    & 4.5   & 1.010(5)   \\
        1.875 & 0.70737(7)    & 5     & 1.030(5) 
  
        \\ \hline\hline
    \end{tabular}
    \label{Kc_exponent}
    \end{threeparttable}
\end{table}

When $\sigma > 2 $, the system undergoes a BKT transition, where the susceptibility $\chi$ is known to follow the scaling behavior: $\chi \sim L^{7/4} (\ln L + C_1)^{1/8} $ at the critical point~\cite{wang2021} where $C_1$ is a nonuniversal constant and needs to be obtained through fitting. Therefore, plotting $\chi L^{-7/4}(\ln L +C_1)^{-1/8}$ as a function of temperature, the curves of different system sizes intersect at the critical point.

\subsection{The Fit of Critical Exponents}
\label{critical_exponent_fit}
 
\begin{figure}[t]
    \centering
    \includegraphics[width=\linewidth]{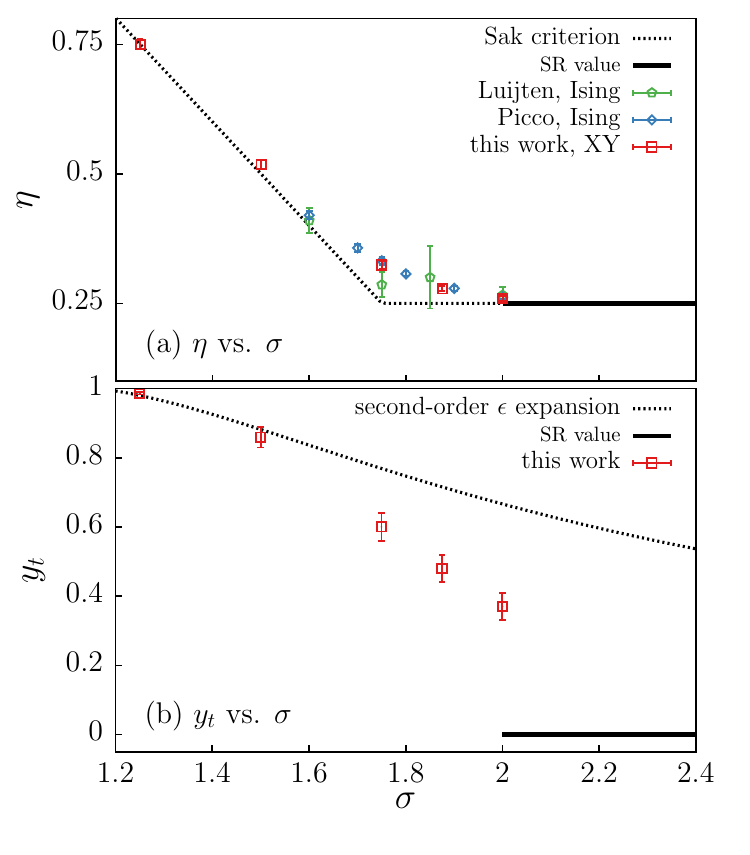}
    \caption{The value of critical exponent at various $\sigma$. (a) The red dots denote the estimates of $\eta$ in this work. Picco and Luijten's results are also contained~\cite{picco2012, Luijten2002}, respectively denoted by blue and green dots. Although the model they studied is the Ising model, our estimated values of $\eta$ are consistent with their results. The dotted lines correspond to Sak's prediction: $\eta=\rm{max}(2-\sigma,\frac14)$. Obviously, our results deviate from Sak's prediction, suggesting that for $1.75<\sigma<2$, the short-range behavior is not recovered. (b) The dotted line represents the results from second-order $\epsilon$ expansion~\cite{Fisher1972} (note that $y_t=1/\nu$). The red dots denote the estimates of $y_t$ in this work. Note that for $\sigma>2$, the LR system undergoes a BKT phase transition instead of a second-order one, with corresponding $y_t = 0$. Therefore, there will be a jump for the value of $y_t$ at $\sigma=2$. This jump is reasonable considering the distinct different type of phase transition for $\sigma=2$ and $\sigma>2$. }
    \label{exponents}
\end{figure}

According to the scaling relationship, only two critical exponents are independent, and here we mainly focus on $\eta$ and the thermal scaling exponent $y_t=1/\nu$, with $\nu$ the correlation-length exponent. In Sak's criterion, $\eta=\max(2-\sigma,\eta_{\rm{SR}})$ where $\eta_{\rm{SR}}=\frac14$ for 2D XY model, so estimating the value of $\eta$ is another way to determine whether $\sigma_*$ equals 1.75 or 2. By fitting the susceptibility $\chi$ and the scaled covariance $K$, the values of $\eta$ and $y_t$ can be numerically estimated. At the critical point, considering the finite-size scaling Eq.~\eqref{chi_scale2} and \eqref{eq:K_scale}, $\chi$ and $K$ are respectively fitted to:
\begin{equation}\label{chiFit}
    \chi=L^{2-\eta}(a+b L^{-\omega})+c
\end{equation}
and
\begin{equation}\label{KFit}
    K=L^{y_t}(a+b\cdot L^{-\omega})+cL^{\eta-2},
\end{equation}
where $b\cdot L^{-\omega}$ refers to the finite-size correction and $c$ derives from the analytic part of freedom energy. During the fitting process, various values of $\omega$ are tested. It is observed that when $\omega$ is excessively large or small, the data cannot be adequately fitted. However, when $\omega$ falls within a certain range, an appropriate fit yields reliable estimates for exponents. The details and results of the fitting are shown in Tables \ref{chi_detail}, \ref{K_detail}, and \ref{Kc_exponent_}, respectively. An interesting point is that our fitted $\eta$ values are consistent with previous results for LR Ising model~\cite{Luijten2002, picco2012}. Figure~\ref{chi_appro_c} and \ref{K_appro_c} visually show the reliability of the fitting.
\begin{table}[b]
    \centering
    \caption{Estimates of two exponents $y_t$ and $\eta$ for nonclassical regime. The results are obtained from the fitting: Eq.~\eqref{chiFit} and \eqref{KFit}. Some of the estimated values of $\eta$ for LR Ising model by Luijten~\cite{Luijten2002} and Picco~\cite{picco2012} are also included for comparison.}
    \begin{threeparttable}
    \begin{tabular}{l|llll}
    \hline\hline
        ~~$\sigma$ & ~~~~$y_t$  & ~~~~$\eta$ & $\eta$ (Luijten) & $\eta$ (Picco) \\ \hline
        1.25     &  0.987(4)  &  0.75(1) && \\
        1.5       &  0.86(3)  &  0.518(8) && \\
        1.75      &  0.60(4)  &  0.324(7) & 0.286(24)& 0.332(8)\\
        1.875     &  0.48(4)  &  0.278(5) && \\
        2         &  0.37(4) &  0.260(5) & 0.266(16)& 0.262(4)  
        \\ \hline\hline
    \end{tabular}
    \label{Kc_exponent_}
    \end{threeparttable}
\end{table}
To be specific, when taking the estimated value of $\eta$ and $y_t$, according to the fitting, $\chi L^{\eta-2}-bL^{-\omega}$ and $KL^{-y_t}-bL^{-\omega}$ should converge to a non-zero value as $L$ increases; otherwise they'll either diverge or reduce to 0. It can be seen that in both figures, the curves corresponding to the estimated values of exponents converge to a plateau, indicating the reliability of the fitting results. Figure~\ref{exponents} presents our numerical estimates of critical exponents at different $\sigma$. Panel (a) displays our estimates for $\eta$ across different $\sigma$ values, alongside the estimates by Picco~\cite{picco2012} and Luijten~\cite{Luijten2002}. 
\begin{table*}[t]
    \centering
    \caption{fitting detail of $\chi$ to Eq.~\eqref{chiFit} for various $\sigma$.}
    \begin{tabular}{p{1.2cm}p{1.2cm}p{2cm}p{2cm}p{2cm}p{2cm}p{1.2cm}p{1.2cm}}
    \hline\hline
        $\sigma$ & $L_\mathrm{min}$ & $\eta$ & $c$ & $a$ & $b$ & $\omega$ &$\chi^2/$DF \\ \hline
        1.25&8 & 0.742(3) & -0.92(6) & 0.64(1) & 1.04(3) & 0.4 & 8.5/10 \\
        &16 & 0.742(4) & -1.0(2) & 0.63(2) & 1.05(5) && 8.5/9 \\ 
        &8 & 0.758(2) & -1.39(9) & 0.74(1) & 1.18(4) & 0.5 & 11.8/10 \\ 
        &16 & 0.755(3) & -1.7(3) & 0.73(2) & 1.24(7) &  & 10.2/9 \\ \hline
        1.5&32 & 0.511(4) & 3.0(7) & 0.50(2) & 0.51(5) & 0.4 & 9.8/8 \\
        &48 & 0.517(5) & 5(2) & 0.52(3) & 0.42(8) && 7.6/7 \\ 
        &16 & 0.525(1) & -1.2(3) & 0.568(6) & 1.22(5) & 0.7 & 10.0/9 \\
        &32 & 0.526(2) & -0.2(10) & 0.575(9) & 1.1(1) && 8.7/8 \\ \hline
        1.75&16 & 0.3169(7) & 1.6(1) & 0.363(2) & 0.535(6) & 0.4 & 7.0/11 \\
        &32 & 0.3164(9) & 1.3(4) & 0.362(3) & 0.541(9) && 6.6/10 \\
        &32 & 0.3305(7) & -3.5(6) & 0.422(2) & 0.91(2) & 0.6 & 10.1/10 \\
        &48 & 0.3298(8) & -5(1) & 0.419(3) & 0.94(3) && 8.4/9 \\ \hline
        1.875& 48 & 0.273(1) & 3.8(7) & 0.386(4) & 0.41(1) & 0.35 & 9.7/10 \\ 
        &64 & 0.274(1) & 5(1) & 0.388(5) & 0.40(1) && 8.8/9 \\
        &16 & 0.2814(5) & 1.0(1) & 0.425(2) & 0.491(6) & 0.45 & 11.7/12 \\ 
        &32 & 0.2812(7) & 0.9(4) & 0.424(2) & 0.494(10) && 11.6/11 \\ \hline
        2&48 & 0.2559(7) & 5.2(6) & 0.465(3) & 0.299(9) & 0.35 & 10.5/10 \\ 
        &64 & 0.2562(8) & 6(1) & 0.466(4) & 0.29(1) && 10.0/9 \\ 
        &32 & 0.2645(2) & -0.6(3) & 0.5131(9) & 0.555(8) & 0.6 & 5.6/11 \\ 
        &48 & 0.2643(3) & -1.2(6) & 0.512(1) & 0.56(1) && 5.0/10 \\ \hline\hline
    \end{tabular}
    \label{chi_detail}
\end{table*}
\begin{table*}[t]
    \centering
    \caption{fitting detail of $K$ to Eq.~\eqref{KFit} for various $\sigma$.}
    \begin{tabular}{p{1.2cm}p{1.2cm}p{2cm}p{2cm}p{2cm}p{1.2cm}p{1.2cm}}
    \hline\hline
        $\sigma$ & $L_\mathrm{min}$ & $y_t$ & $a$ & $b$ & $\omega$ &$\chi^2/$DF \\ \hline
        1.25& 8 & 0.991(2) & 0.444(5) & 0.067(8) & 0.4 & 8.2/11 \\ 
        &16 & 0.989(2) & 0.453(7) & 0.05(1) && 6.2/10 \\
        &8 & 0.9833(6) & 0.473(1) & 0.56(7) & 2 & 8.4/11 \\
        &16 & 0.9839(7) & 0.472(2) & 0.9(3) && 7.4/10 \\ \hline
        1.5&8 & 0.830(2) & 1.60(3) & -1.21(4) &0.1& 9.3/10 \\ 
        &16 & 0.826(3) & 1.67(5) & -1.29(6) && 7.4/9 \\ 
        &16 & 0.886(2) & 0.69(1) & -0.43(3) &0.5& 10.8/9 \\ 
        &32 & 0.883(3) & 0.71(2) & -0.49(6) & &9.3/8 \\ \hline
        1.75&32 & 0.558(4) & 5.1(2) & -5.1(2) &0.1& 11.8/11 \\ 
        &48 & 0.553(5) & 5.4(3) & -5.5(3) && 9.5/10 \\ 
        &48 & 0.640(3) & 1.48(3) & -3.3(2) &0.6& 9.4/10 \\
        &64 & 0.640(4) & 1.48(4) & -3.3(3) & &9.4/9 \\ \hline
        1.875&16 & 0.441(2) & 6.9(1) & -7.1(1) &0.1& 10.8/12 \\
        &32 & 0.439(3) & 7.0(2) & -7.3(3) & &10.2/11 \\ 
        &48 & 0.518(3) & 2.04(4) & -3.9(2) &0.5& 8.6/10 \\ 
        &64 & 0.518(4) & 2.04(6) & -3.9(3) && 8.6/9 \\ \hline
        2&32 & 0.337(3) & 8.9(3) & -9.6(3) &0.1& 9.7/11 \\
        &48 & 0.332(4) & 9.3(4) & -10.2(5) & &7.6/10 \\ 
        &64 & 0.416(4) & 2.53(7) & -5.6(3) & 0.5& 6.9/9 \\
        &128 & 0.411(6) & 2.6(1) & -6.3(8) & &5.4/7 \\ \hline
    \end{tabular}
    \label{K_detail}
\end{table*}
Despite the difference in models (Luijten and Picco studied the 2D LR Ising model, whereas we are studying the 2D LR XY model), our results are consistent with Picco's, which does not agree with Sak's prediction. In our results, $\eta$ interpolates smoothly between $\eta=2-\sigma$ for $\sigma$ near 1 and $\eta=\eta_{\text{SR}}=\frac14$ for $\sigma>2$. For $\sigma<2$, $\eta$ remains significantly higher than $\frac14$, indicating the phase transition doesn't belong to the short-range universality class. For $\sigma=2$, the estimated value of $\eta$ still slightly exceeds $\frac14$, which could be attributed to some logarithmic corrections in the behavior of $\chi$ at the marginal point $\sigma=2$. Panel (b) in Fig.~\ref{exponents} demonstrates our numerical estimates of $y_t$, as well as the $\epsilon$ expansion results up to the second order~\cite{Fisher1972}.  {It can be seen that our numerical data starts to deviate from the $\epsilon$ expansion result from $\sigma\approx1.5$. This should also apply to the value of $\eta$ according to the panel (a). Therefore, the $\epsilon$ expansion result near $\sigma=1$, $\eta=2-\sigma$ should not always hold along the whole nonclassical regime, especially for $\sigma\geq 1.5$ cases. However, Sak criterion simply adopt the relation for $\sigma\leq1.75$, which we considered unreliable.}

\subsection{Subleading Magnetic Exponents}
\label{double_power}

\begin{figure*}
    \centering
    \includegraphics[width=0.95\linewidth]{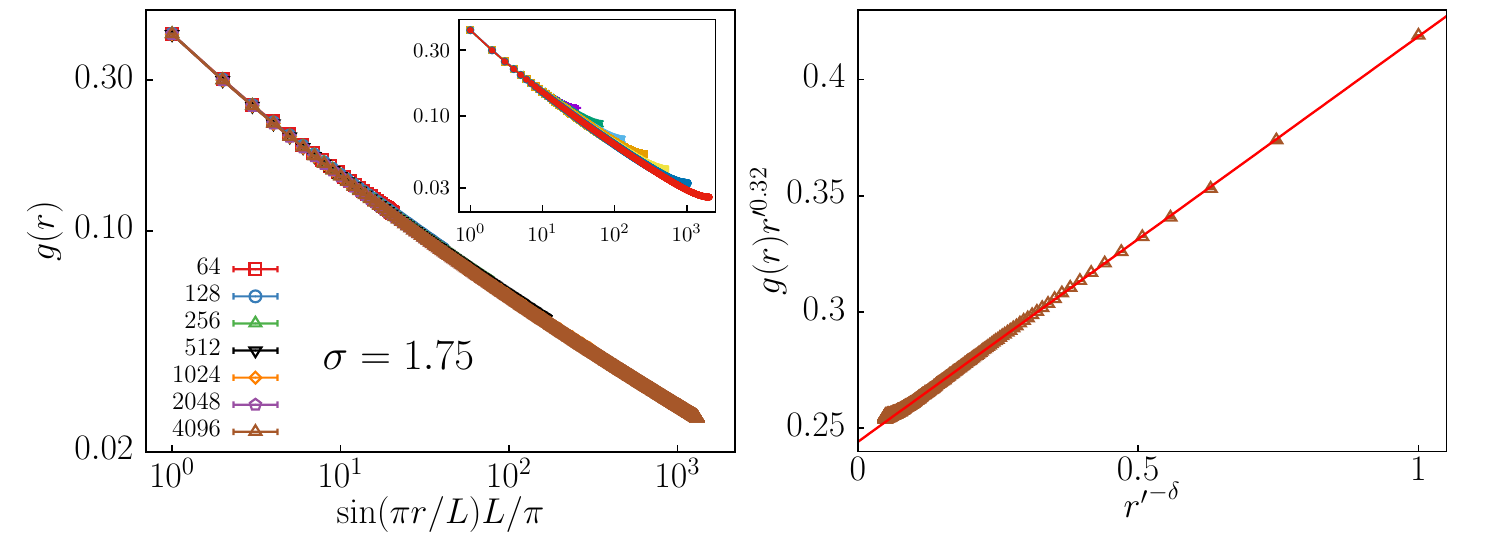}
    \caption{(Left) The correlation function $g(r)$ for systems of different sizes is plotted for $\sigma=1.75$. Both the main figure and inset are shown in double-logarithmic coordinates for convenient observation of the power-law behavior. In the inset, with the x-axis representing $r$, the correlation function curves exhibit a strong boundary effect, i.e., a noticeable tendency to bend upwards as $r$ approaches $L/2$. In the main figure, the x-axis is replaced by $\sin(\pi r/L)L/\pi$ to reduce this boundary effect. Furthermore, in this representation, the behavior of the correlation function at short distances remains unaffected.
    (Right) Rescaled correlation function $g(r)r'^{\eta}$ versus ${r'}^{-\delta}$ for $L = 4096$, with $\eta=0.32$ and $\delta=0.42$. The red line represents the fitted line. This figure visually demonstrates the contribution of the subleading magnetic exponent to $g(r)$ (discussed in the text).}
    \label{corrFunc}
\end{figure*}

\begin{figure}
    \centering
    \includegraphics[width=\linewidth]{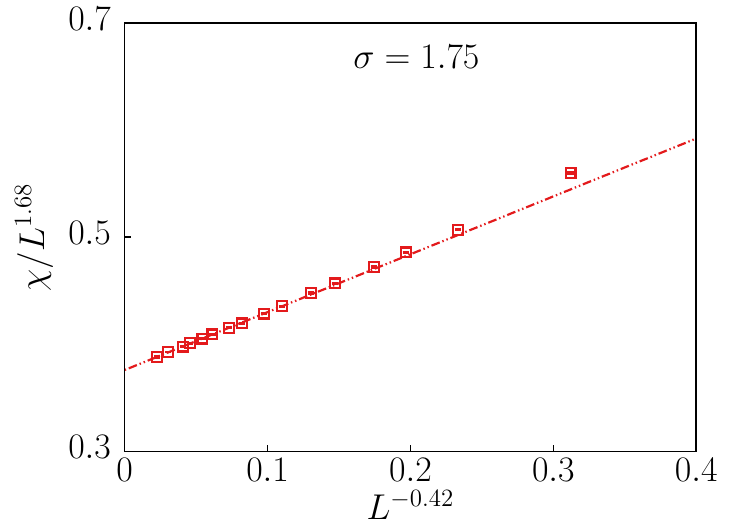}
    \caption{The plot of $\chi/L^{2-\eta}$ versus $L^{-\delta}$ for $\sigma = 1.75$, where $\eta=0.32$ and the corresponding $\delta=0.42$, obtained from the fitting of Eq.~\eqref{corr_fit}. As the system size \( L \rightarrow \infty \), the red dots approach a straight line, indicating the scaling behavior: $\chi\sim L^{2-\eta}(a+b\cdot L^{-\delta})$ with $\eta=0.32$ and $\delta=0.42$. This shows the consistency between the fitting of $\chi$ and $g(r)$.} 
    \label{demon}
\end{figure}

 {As mentioned before, Ref.~\cite{angelini2014} suggests that Picco might underestimate the finite-size correction, which finally caused the overestimate of $\eta$. By studying the short-distance correction to the correlation function, they proposed that when properly considering the finite-size correction term, $\eta$ would revert to $0.25$ at $\sigma=1.75$, consistent with Sak's criterion.} In this subsection, following the same routine in Ref.~\cite{angelini2014}, we also study the correlation function at the critical point for the 2D LR XY model. In particular, we focus on the case of $\sigma=1.75$ where $\eta$ is estimated to be $0.324(7)$, and similar procedures can apply to other $\sigma$. Our estimated value of $\eta$ is taken to be $0.32$, with the corresponding correction exponent $\omega$ in Eq.~\eqref{chiFit} being $0.45$. By carefully fitting the correlation function, we identify a subleading power-law correction likely arising from the subleading magnetic exponent $y_{h,2}$ of the transition~\cite{blote1995}. The fitting results are compatible with our estimates of $\eta$, thus validating our previous analysis.

The two-point correlation function is defined as $g(r)=\langle \boldsymbol{S}(0)\cdot\boldsymbol{S}(\boldsymbol{r})\rangle$, and at the critical point, it asymptotically decays as a power of the separation, $g(r)\propto r^{-\eta}$ as $r \rightarrow \infty$. However, one would expect additional corrections to $g(r)$ that decay faster than the leading term. These corrections are, in general, difficult to observe in SR models. It is argued in Ref.~\cite{angelini2014} that $g(r)$ in the LR Ising model at the criticality is controlled by two power-law decays, as $g(r)=r^{-\eta}(a+br^{-\delta})$. The term $r^{-\delta}$ causes faster decay in short distances and reduces to 0 at large $r$ when the correlation function gradually returns to the single power scaling, i.e., $~r^{-\eta}$. Such correction, if not properly considered, could lead to an unreliable estimation of $\eta$~\cite{angelini2014}.

We measure the correlation function in horizontal and vertical axes:
\begin{equation}
    g(r)=\frac{1}{2N_{s}}\sum_{\langle i,j\rangle\in \{N_s\}}{\left(\langle \boldsymbol{S}_{i,j}\cdot\boldsymbol{S}_{i+r,j}\rangle+\langle \boldsymbol{S}_{i,j}\cdot\boldsymbol{S}_{i,j+r}\rangle\right)},
\end{equation}
where the summation only goes through $N_s$ randomly chosen sites instead of the entirety, and $N_s=100$ in our practice. In this way, the complexity of this measurement remains $\mathcal{O}(N)$ instead of $\mathcal{O}(N^2)$, which allows us to measure the system with a larger size and simulate more samples. The cost is the slightly higher uncertainty in the measurement, but this can be compensated by involving more samples. 

\begin{figure*}[ht]
\centering
\includegraphics[width=1.0\linewidth]{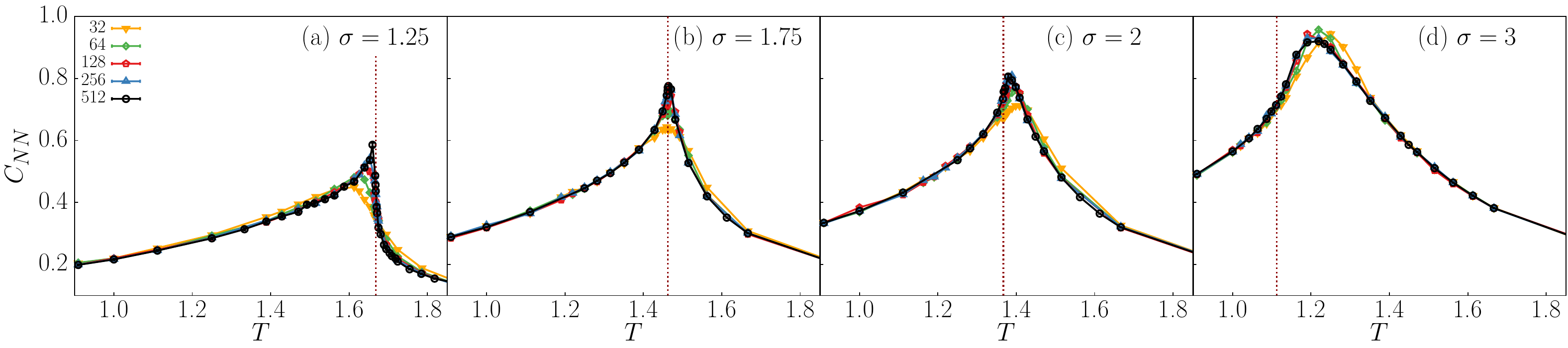}	
\caption{The plot of $C_{\text{NN}}$ versus $T$ is shown for (a) $\sigma = 1.25$, (b) $\sigma = 1.75$, (c) $\sigma = 2$, and (d) $\sigma = 3$. The dashed vertical lines indicate the transition points for various $\sigma$ values. For $\sigma \leq 2$, finite yet sharp peaks are observed at the critical points, consistent with our prediction of second-order phase transitions. In contrast, the smooth and broad peak for $\sigma = 3$ is the typical behavior of a BKT transition.}
\label{fig:Cv}
\end{figure*}

 {The inset of the left panel in Fig.~\ref{corrFunc} demonstrates the correlation function with various system size $L$ where the double-log scale is used to observe the power-law behavior. The flattening tails of the correlation functions are due to the finite-size effect. To diminish such effect for a better observation of the power-law behavior, we follow the treatment of Ref.~\cite{angelini2014} to choose the x-axis as $\sin(\pi r/L)L/\pi$ in the main figure. It does not affect the correlation function in short distances since $\sin(\pi r/L)L/\pi\approx (\pi r/L)L/\pi=r$. Moreover, as can be seen in the main figure, it drastically reduces the boundary effect.} In  the main figure, $g(r)$ exhibits a fast decay in the short distance and then decays slowly as $r$ increases, which suggests an additional correction to the correlation function. We then perform a detailed fit to the correlation function of the largest system size $L=4096$ with $\eta$ fixed at $0.32$:
\begin{equation}
    g(r)={r'}^{-\eta}(a+b{r'}^{-\delta}),
\label{corr_fit}
\end{equation}
where $r'=\sin(\pi r/L)L/\pi$. We use data ranging from $r=1$ to $600$ for fitting, minimizing the impact of boundary effects. The fit results in a corresponding $\delta$ value of $0.42$. The right panel in Fig.~\ref{corrFunc} visually shows the quality of fitting, where the x and y-axis are respectively ${r'}^{-\delta}$ and $g(r){r'}^{\eta}$, and thus the data points should approximately form a straight line according to Eq.~\eqref{corr_fit}. As shown in the figure, the data points fit well with a straight line and are only slightly flattened at large distances due to the boundary effect. Therefore, the fitting of the correlation function is compatible with $\eta=0.32$.

We also demonstrate that the correction to correlation has been properly included in the fitting of $\chi$. It is argued in Ref.~\cite{angelini2014} that Picco did not sufficiently account for the finite-size corrections deriving from the second term in the double power of the correlation function, leading to an overestimation of $\eta$. Hence, we compare the value of correction exponent $\omega$ considered in the fitting of $\chi$ using Eq.~\eqref{chiFit}, and of the second exponent $\delta$ in Eq.~\eqref{corr_fit}. For $\eta = 0.32$, the corresponding $\omega$ in the fit of Eq.~\eqref{chiFit} is approximately $0.45$, while the fit of Eq.~\eqref{corr_fit} gives a $\delta$ of $0.42$. Since these values are very close, we believe that the correction term $bL^{-\omega}$ in Eq.~\eqref{chiFit} has already accounted for the subleading term $r'^{-\delta}$ of the correlation function. Figure.~\ref{demon} plots $\chi/L^{2-\eta}$ as a function of $L^{-\delta}$ with $\eta=0.32$ and $\delta=0.42$, where the data points asymptotically follows a linear behavior. This suggests that $\chi$ indeed scales as $\sim L^{2-\eta}(a+b\cdot L^{-\delta})$ and our previous estimate of $\eta$ by fitting $\chi$ is reliable.

The additional power-law correction to $g(r)$ should originate from a subleading relevant magnetic scaling field. In the context of RG, the free-energy density depends on not only the leading thermal and magnetic scaling fields $t_1 \equiv t$, $h_1 \equiv h$ but rather a full set of scaling fields. Each scaling field is associated with a particular eigenvalue of the RG transformation. In particular, the subleading thermal and magnetic scaling fields $t_2$ and $h_2$ correspond to the subleading eigenvalues in the even sector and odd sector, respectively. The contributions of these subleading exponents are very difficult to observe explicitly from numerical simulations because they either decay very fast or have very small, even zero amplitude. For instance, for the NN Ising model, one has $y_{t,2}=-4/3$ and $y_{h,2} = 13/24 \approx 0.542$~\cite{domb1987}. The subleading thermal field $t_2$ is irrelevant, whose contribution to FSS decays quickly. As for the subleading magnetic field $h_2$, despite being relevant, it is commonly believed to be redundant~\cite{blote1989,baillie1992}. Recently, the exponent $y_{h,2}$ has been observed in the Fortuin-Kasteleyn geometric representation of the Ising model~\cite{e27040418}.

Taking into account the dependence on the additional magnetic scaling field $h_2$, the singular part of free energy density for a finite system near criticality scales as $f\sim L^{-d} f (t_1 L^{y_{t,1}},h_1 L^{y_{h,1}},h_2 L^{y_{h,2}}, \dots)$, where $y_{t,1} \equiv y_t$, $y_{h,1} \equiv y_h$, and $y_{h,2}$ are the corresponding scaling exponents. Consequently, at critical point $t=0$, the correlation function $g(r)$, acquires additional power-law terms as $g(r)\sim g_0 r^{2y_{h,1}-2d} + g_1 r^{y_{h,1}+y_{h,2}-2d} + g_2 r^{2y_{h,2}-2d}$ ~\cite{blote1995}. From our fitting, the anomalous dimension of the subleading power-law decay is $\eta' \equiv \eta + \delta = 0.74$. Therefore, if this exponent is from the $r^{y_{h,1}+y_{h,2}-2d}$ term, one has $y_{h,2} = 2- (\eta'-\eta/2) \approx 1.42$; or, if it is from $r^{2y_{h,2}-2d}$, then $y_{h,2} = 2-\eta'/2 \approx 1.63$. Either case is possible, though, from a symmetry consideration, we believe the latter case is more likely. Nonetheless, one could conclude that the correction to $g(r)$ is most likely due to $y_{h,2}$.

We are glad to observe the contribution of the subleading magnetic field at criticality in the LR XY model, which is difficult to identify numerically in previous studies of SR cases. In our opinion, this strong correction to $g(r)$ is another piece of evidence that the system is not in the SR universality class at $\sigma = 1.75$, since the two cases do not have the same subleading magnetic exponent. Note that the double power correlation function observed in the LR Ising model could also be attributed to the contribution of $y_{h,2}$, suggesting a different universality class with the SR one at $\sigma = 1.75$.

\subsection{The properties of the specific-heat-like quantity}
\label{Cv}

 {As a supplemental indication for the crossover from SR universality to LR universality, the specific heat $C$ exhibits somewhat different behaviors in the two regimes. In the SR regime, the specific heat $C$ exhibits a smooth and broad peak, which is a hallmark of the BKT transition. Conversely, as the system approaches a second-order phase transition, the singular part of $C$ scales $\sim \left|t\right|^{-\alpha}$, where $t$ is the reduced temperature and $\alpha$ is the specific-heat critical exponent. 
When $\alpha > 0$, the specific heat diverges at the critical temperature $T_c$. However, for $\alpha < 0$, $C$ remains finite at $T_c$ despite the presence of critical singularities. In such cases, the specific heat typically develops a sharp peak or even a nonanalytic kink at the transition point, which can serve as a hint of a second-order phase transition.

Directly measuring the full energy density and specific heat for long-range interacting systems is computationally expensive. Hence, we compute the energy of nearest neighbors for simplicity, defined as
\begin{align}
    \varepsilon  = L^{-2}\sum_{\langle ij \rangle} \bm{S}_{i} \cdot \bm{S}_{j},
\end{align}
where the summation is over all NN pairs. This energy-like quantity shares the same critical scaling form of the energy density $\langle E \rangle$. Thus, it is natural to define a specific-heat-like quantity that possesses the identical scaling form of the specific heat,
\begin{equation}
    C_{\text{NN}} = \beta^2 L^2 \left( \langle \varepsilon^2\rangle - \langle \varepsilon  \rangle^2 \right),
\end{equation}

In Fig.~\ref{fig:Cv}, we plot $C_{\text{NN}}$ versus $T$ for $\sigma = 1.25$, $1.75$, $2$, and $3$. For $\sigma = 3$, $C_{\text{NN}}$ exhibits a smooth and broad peak slightly above the transition temperature $T_c$, which is a typical feature of the BKT transition. In contrast, for $\sigma = 1.25$, $1.75$, $2$, finite yet sharp peaks are clearly observed near $T_c$. As the $\sigma$ decreases, the peak becomes sharper and more asymmetric, indicating nonanalytic behavior in the free energy. This observation is consistent with our analysis of critical exponents. For a system below its upper critical dimension, the specific-heat critical exponent $\alpha$ is related to the correlation length exponent $\nu$ via the hyperscaling relation $\alpha=2-d\nu$. Table~\ref{Kc_exponent_} shows that $y_t < 1$ for  $\sigma = 1.25$, $1.75$, $2$, which implies $1/y_t =\nu > 1$ and $\alpha < 0$. Therefore, at these $\sigma$ values, the specific heats develop cusps at the critical points.

Moreover, since much stronger evidence has already been provided from the low-T properties and the power-law divergence of $\xi$ as the critical point is approached from the high-T side, a comprehensive analysis of $C_{\text{NN}}$ is beyond the scope of this work. We note that, even though $C_{\text{NN}}$ alone does not provide definitive evidence for locating the crossover point $\sigma_*$, it offers valuable supplementary insight. The distinct behaviors observed in the $\sigma \le 2$ and $\sigma>2$ regimes are consistent with our previous analysis and provide additional support for the existence of a crossover at $\sigma =2$.}

\section{Conclusion}\label{sec:conclusion}
In this study, we provide a comprehensive understanding of the two-dimensional long-range (LR) XY model with algebraically decaying interactions. We perform large-scale Monte Carlo simulations to study the low-T, high-T, and critical properties of the model at various $\sigma$. Our results provide compelling evidence that the threshold between long-range and short-range universality lies at $\sigma_* = 2$, in contrast to Sak's scenario~\cite{sak1973} and previous theoretical predictions~\cite{Giachetti2021, Giachetti2022}. The main findings are summarized below:
\begin{enumerate}
\item For $\sigma \leq 2$, the system is shown to exhibit long-range order and Goldstone mode excitations in the low-T phase, while for $\sigma > 2$, the system exhibits quasi-long-range order as in the NN case.
\item We reveal the distinct growth behavior of the correlation length $\xi$ in the high-T phase as the system approaches $T_c$. For $\sigma \leq 2$, the correlation length follows a power-law divergence, the signature of a second-order transition; for $\sigma > 2$, the typical exponential growth of the BKT transition is observed.
\item The critical points and critical exponents of the model in the regime $1 < \sigma \leq 2$ are determined. The estimates of $y_t$ and $\eta$ start to deviate from the $\epsilon$-expansion result from approximately $\sigma=1.50$, and the deviation becomes bigger as $\sigma$ further increases. This strongly suggests that the $\epsilon$-expansion results, like $\eta=2-\sigma$, are perturbative and should not be regarded as quantitatively true, particularly when $\sigma$ approaches 2. Moreover, our estimated values of critical exponents show distinct difference for $\sigma\leq2$ and $\sigma>2$.
\end{enumerate}

 {Another important observation from our study is the clear demonstration of a LRO phase at $\sigma =2$, which is in contrast with the theoretical scenario proposed in Ref.~\cite{PhysRevLett.87.137203}. In that work, the author employed the Bogoliubov inequality in combination with a \textit{reductio ad absurdum} argument, stating that if $\int_{\text{BZ}} 1/\tilde{E} (\bm{k}) d\bm{k} = +\infty$, where $\tilde{E}(\bm{k}) = \sum_{\bm{r}}J(\bm{r})(1-\exp(i\bm{r}\cdot\bm{k}))$, then the magnetization must vanish in the thermodynamic limit, i.e., $\lim_{|B|\to0}\langle S^z \rangle = 0$ for any finite $T$. This condition implies that 2D LR Heisenberg and XY models with algebraic decaying coupling cannot have finite temperature spontaneous magnetization for $\sigma\ge2$. However, we have presented a series of strong numerical evidence that the marginal case $\sigma=2$ does indeed support a LRO phase.
It is also worth emphasizing that the original Mermin-Wagner theorem excluded the spontaneous breaking of continuous symmetry when the second moment of the interaction is finite, i.e., $\sum_{\bm r}{\bm{r}}^2 J(\bm{r}) < \infty$~\cite{mermin1966}. For the interaction $J(r) \propto 1/r^{2+\sigma}$ in 2D, this condition fails at $\sigma = 2$, where the summation diverges logarithmically. 
In any case, the discrepancy between our numerical results and existing theoretical arguments underscores that the nature of the marginal case at $\sigma=2$ remains an open and subtle question, deserving further theoretical and numerical investigation.}

\acknowledgments
We acknowledge the support by the National Natural Science Foundation of China (NSFC) under Grant No. 12204173 and No. 12275263, as well as the Innovation Program for Quantum Science and Technology (under Grant No. 2021ZD0301900). YD is also supported by the Natural Science Foundation of Fujian Province 802 of China (Grant No. 2023J02032).
\\
\appendix
\section*{Appendix A: Extrapolation Detail}
\begin{table*}[!ht]
    \centering
    \caption{Specific values of parameters obtained from the fit of Eq.~\eqref{extra_fit} for $n=9$ and different $(L_{\mathrm{min1}},L_{\mathrm{min2}})$.}
    \label{extra_fit2_detail}
    \begin{tabularx}{\textwidth}{XXXXXXXXX}
    \hline\hline
$L_\text{min}$ & $a_1(\times10^0)$ & $a_2(\times10^0)$ & $a_3(\times10^2)$ & $a_4(\times10^3)$ & $a_5(\times10^3)$ & $b$ & $\omega$ & $\chi^2$/DF \\ \hline
        (128, 256) & 2.22(18) & -8.2(49) & 2.83(51) & -1.30(23) & 1.63(36) & 1.44(13) & 0.538(19) & 191.1/93 \\ 
        (128, 512) & 2.24(18) & -12.2(48) & 3.31(49) & -1.51(21) & 1.96(33) & 1.37(10) & 0.514(16) & 91.6/71 \\ 
        (256, 256) & 2.06(19) & -2.7(51) & 2.15(55) & -0.97(25) & 1.10(39) & 1.12(15) & 0.494(25) & 148.3/80 \\
        (256, 512) & 2.14(17) & -7.4(47) & 2.70(48) & -1.22(21) & 1.49(33) & 1.30(16) & 0.505(23) & 55.9/58 \\ 
        (512, 256) & 2.07(20) & -3.4(58) & 2.20(63) & -0.98(29) & 1.09(47) & 1.01(15) & 0.476(27) & 141.6/73  \\ 
        (512, 512) & 2.11(20) & -6.4(55) & 2.57(59) & -1.16(27) & 1.39(42) & 1.23(24) & 0.496(34) & 54.9/51 \\ 
        \hline\hline
    \end{tabularx}
\end{table*}

\begin{table*}[!ht]
    \centering
    \caption{Specific values of parameters obtained from the fit of Eq.~\eqref{extra_fit} for $n=9$ and different $(L_{\mathrm{min1}},L_{\mathrm{min2}})$.}
    \label{extra_fit_detail}
    \begin{tabularx}{\textwidth}{XXXXXXXXX}
    \hline\hline
$L_\text{min}$ & $a_1(\times10^0)$ & $a_2(\times10^1)$ & $a_3(\times10^3)$ & $a_4(\times10^4)$ & $a_5(\times10^5)$ & $a_5(\times10^5)$ & $a_5(\times10^5)$ & $\chi^2$/DF \\ \hline
        (64, 256) & 2.14(33) & -5.3(20) & 1.57(44) & -1.79(47) & 1.07(26) & -3.17(71) & 3.64(77) & 113.2/50 \\
        (64, 512) & 2.26(27) & -6.1(16) & 1.78(37) & -2.03(40) & 1.20(22) & -3.49(60) & 3.94(66) & 68.1/46 \\
        (128, 256) & 1.66(35) & -2.3(22) & 0.90(48) & -1.07(52) & 0.67(28) & -2.11(78) & 2.53(84) & 79.3/41 \\
        (128, 512) & 1.82(24) & -3.5(15) & 1.19(34) & -1.40(36) & 0.86(20) & -2.62(55) & 3.05(60) & 33.0/37 \\
        (256, 256) & 1.62(55) & -1.7(32) & 0.69(67) & -0.80(69) & 0.51(37) & -1.64(99) & 2.0(10) & 63.6/29 \\
        (256, 512) & 1.85(33) & -3.3(19) & 1.06(41) & -1.21(42) & 0.74(23) & -2.25(61) & 2.64(64) & 19.1/25 \\
        \hline\hline
    \end{tabularx}
\end{table*}

\label{appen_A}
As mentioned in the main text, using the FSS ansatz~\eqref{extrapolation_eq}, we extrapolate the correlation length $\xi$ to its thermodynamic-limit value. For $\sigma=2$ and $3$, the ansatz~\eqref{extrapolation_eq} is fitted to Eq.~\eqref{extra_fit} and \eqref{extra_fit2} respectively. To further account for the finite-size correction, 
we gradually increase the minimum system size $L_\text{min}$ included in the fitting. It is found that for both cases, finite-size corrections are more pronounced in the range $\xi/L \in (0,0.35)\cup(0.65,1)$ than in other intervals. Therefore, two different minimum system sizes $L_\text{min1}$ and $L_\text{min2}$ are selected for two separate intervals, namely the range $\xi/L \in (0.35,0.65)$ and $\xi/L \in (0,0.35)\cup(0.65,1)$ respectively, with
$L_\text{min1} \leq L_\text{min2}$. Fitting results with different $(L_\text{min1},L_\text{min2})$ for $\sigma=2$ and $3$ are presented in Table~\ref{extra_fit2_detail} and \ref{extra_fit_detail} respectively. The extrapolation results in Fig.~\ref{extrapolation} use the optimal fit with $(L_\text{min1}=256,L_\text{min2}=512)$ for $\sigma=2$ and $(L_\text{min1}=128,L_\text{min2}=512)$ for $\sigma=3$. Further extrapolation details can be found in Ref.~\cite{Yao2025}

\bibliography{long_range_xy_model}

\end{document}